%% file: main.tex
\documentclass{stsci_report}

\usepackage{multirow}
\usepackage{array}
\usepackage{graphicx}
\usepackage{siunitx}
\usepackage{todonotes}
\usepackage{pdfpages}
\usepackage{longtable}
\usepackage{booktabs}
\usepackage{float}
\usepackage{longtable}
\usepackage{indentfirst}
\usepackage{caption}
\usepackage{hyperref}
\usepackage{multirow}
\usepackage{paracol}
\usepackage{colortbl}
\usepackage{setspace} 
\usepackage{amsmath}
\usepackage{enumitem}
\usepackage{soul}
\usepackage{xcolor}
\usepackage{hyperref}
\usepackage{array}
\hypersetup{
    colorlinks=true,
    linkcolor=black,
    urlcolor=blue,
    citecolor=black
}
\copyrighttext{Copyright\copyright\ \the\year\ The Association of Universities for Research in Astronomy, Inc. All Rights Reserved.}

\presubtitle{Instrument Science Report WFC3 2025-09}
\title{Testing and Validation of the Updated Pixel-Based Non-Linearity Calibration \\ File for WFC3/IR}
\author{K.\ Huynh, V.\ Bajaj, M.\ Marinelli, J.\ Mack, S.\ Shenoy, N.\ Grogin}
\date{November 4, 2025 \\ Last Updated: February 4, 2026}
\begin{document}

\maketitle
\abstract{The WFC3/IR channel has an innate non-linear response to incident photons, which is corrected for in the \texttt{calwf3} pipeline with the \texttt{NLINFILE} reference file. The 2009 solution is based on an average polynomial correction for each IR quadrant and is found to be poorly constrained at high fluence levels ($e^-$) approaching the saturation limit. Using a variety of image types, sources, and sample sequences, we test a new pixel-based linearity correction developed by \textcite{shenoy2025}. In nearly all cases, the new correction improves the linearity at fluence levels higher than $\sim$50,000 $e^-$, with improvements up to 7\% for pixels with fluences approaching the saturation limit ($\sim$80,000 $e^- $) in the last \texttt{ima} reads. The pixel-based solution also significantly decreases the number of cosmic rays erroneously flagged (due to non-linearity correction errors) during ramp fitting in \texttt{calwf3}, leading to improved photometric accuracy in the calibrated \texttt{flt} data and higher signal-to-noise ratios, particularly in Quad 1 (upper-left detector quadrant). Because the new solution tends to make sources brighter, we recalibrate the five HST flux standards used to compute the IR zeropoints and find a negligible impact ($\sim$0.1--0.2\%) on the published values by \textcite{Calamida2024}, smaller than the RMS dispersion ($\sim$0.5\%) in the observed to synthetic flux ratios for all five flux standards. The new NLINFILE \texttt{9au15283i\_lin.fits} was delivered to CRDS in October 2025 and will be used to reprocess all WFC3/IR imaging and grism observations in the MAST archive.\\ 
\textbf{An updated reference file \texttt{a2412448i\_lin.fits} was delivered in February 2026, improving the results at the highest fluence levels by a few tenths of a percent. Please consult the Addendum for details.}
}

\newpage
\section*{Introduction}
The Wide Field Camera 3 Infrared (WFC3/IR) channel is a HgCdTe detector with an intrinsically non-linear response to incident photons \parencite{2009Hilbert}. Pixel counts that are below the saturation threshold of the IR detector and whose measured flux deviates from linearity are currently modeled by a third-order polynomial, which relates the measured and idealized signal. The WFC3 data calibration pipeline, \texttt{calwf3}, corrects for this non-linearity using a calibration reference file called the \texttt{NLINFILE}. This \texttt{NLINFILE} contains the coefficients of a third-order polynomial applied to each pixel on the detector to correct for the non-linear response. The structure and format of the \texttt{NLINFILE} is described in more detail in \textcite{shenoy2025}.
\bigbreak
The 2009 \texttt{NLINFILE} (\texttt{u1k1727mi\_lin.fits}, hereafter referred to as the current correction) uses a set of quadrant-averaged polynomials derived from Tungsten lamp flats acquired on the ground during TV3 testing \parencite{2009Hilbert}. These flats were acquired using the \texttt{STEP50} sample sequence, in which the signal is sampled more frequently in the early reads and then at uniform 50-second intervals in the later reads\footnote{For more information on the predefined sample sequence modes for the WFC3/IR channel, please see the \href{https://hst-docs.stsci.edu/hpiom}{HST Phase II Proposal Instruction Handbook.}}. This sampling pattern makes it difficult to constrain the non-linearity for pixels at high fluence levels approaching the full well limit. A new pixel-based non-linearity correction file has been derived using on-orbit \texttt{SPARS25} internal flat fields from WFC3 calibration (CAL) programs designed to monitor the IR linearity. WFC3 ISR 2025--08 \parencite{shenoy2025} describes the methodology used to derive the new coefficients and generate the new pixel-based correction file. The aforementioned report also includes a map of the IR full saturation limit for each pixel by \textcite{2009Hilbert}, populated in the \texttt{NODE} extension of the \texttt{NLINFILE}. The values in this extension (i.e. saturation levels) are unchanged from the original analysis based on TV3 data.
\bigbreak
In this report, we highlight improvements in the new pixel-based non-linearity correction file compared with the quadrant-based solution. We compare the results from testing on various datasets, including internal flat fields, star clusters, and CALSPEC standard stars (imaging and grisms), which are used for HST flux calibration. 
Finally, we measure changes in the observed count rates for the five flux standards in all 15 IR filters in order to check the absolute flux calibration (zeropoints) computed by \textcite{Calamida2024}.

\section*{Internal Flat Fields}
\begin{figure}[H]
\begin{center}
\includegraphics[width=.88\linewidth]{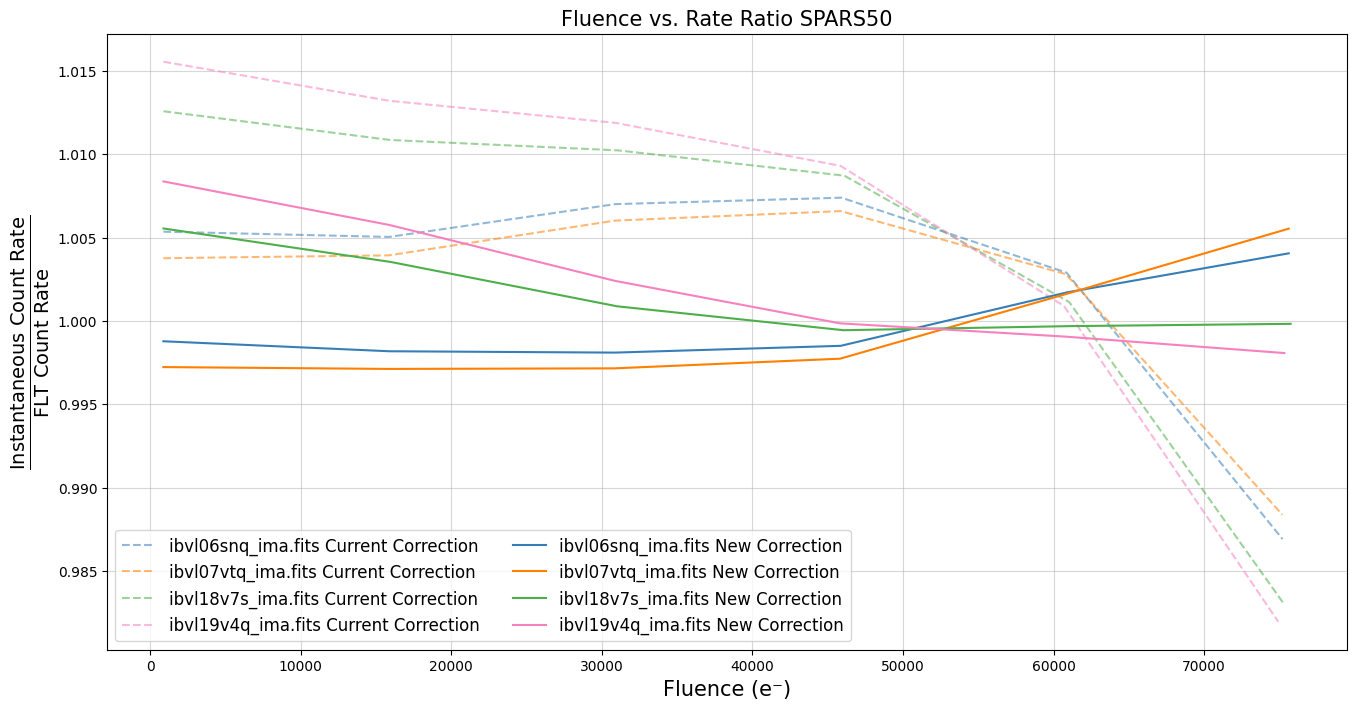}
\end{center}
\caption{\textit{Mean ratio between the \texttt{ima} instantaneous count rate and the \texttt{flt} count rate vs. the mean fluence level ($e^-$) for a set of INTFLATs taken in \texttt{SPARS50} sample sequence. Dashed lines are INTFLATS calibrated with the 2009 \texttt{NLINFILE}, while solid lines show ramps calibrated with the new \texttt{NLINFILE}. The last four reads are saturated and excluded from the plot. The new correction reduces the peak-to-peak range from $\sim3\%$ to $\sim1\%$ across all fluence levels.}
}
\label{fig:intflat_spars50}
\end{figure}

\begin{figure}[H]
\begin{center}
\includegraphics[width=.88\linewidth]{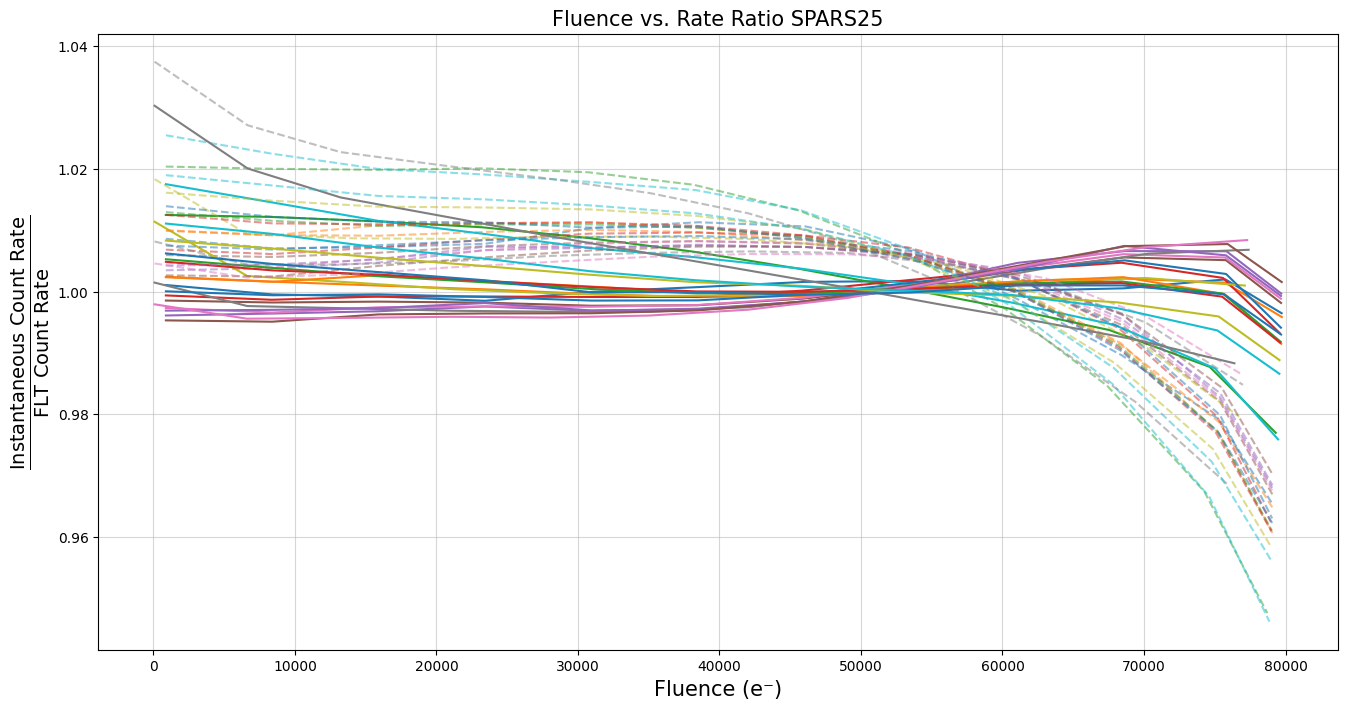}

\end{center}
\caption{\textit{Same as Figure \ref{fig:intflat_spars50}, but for a set of \texttt{SPARS25} INTFLATS. The last three reads are saturated and excluded from the plot. The new correction reduces the peak-to-peak range from $\sim$6--7$\%$ to $\sim2\%$, representing the largest improvement observed across all sample sequences tested.}
}
\label{fig:intflat_spars25}
\end{figure}

\begin{figure}[H]
\begin{center}
\includegraphics[width=.90\linewidth]{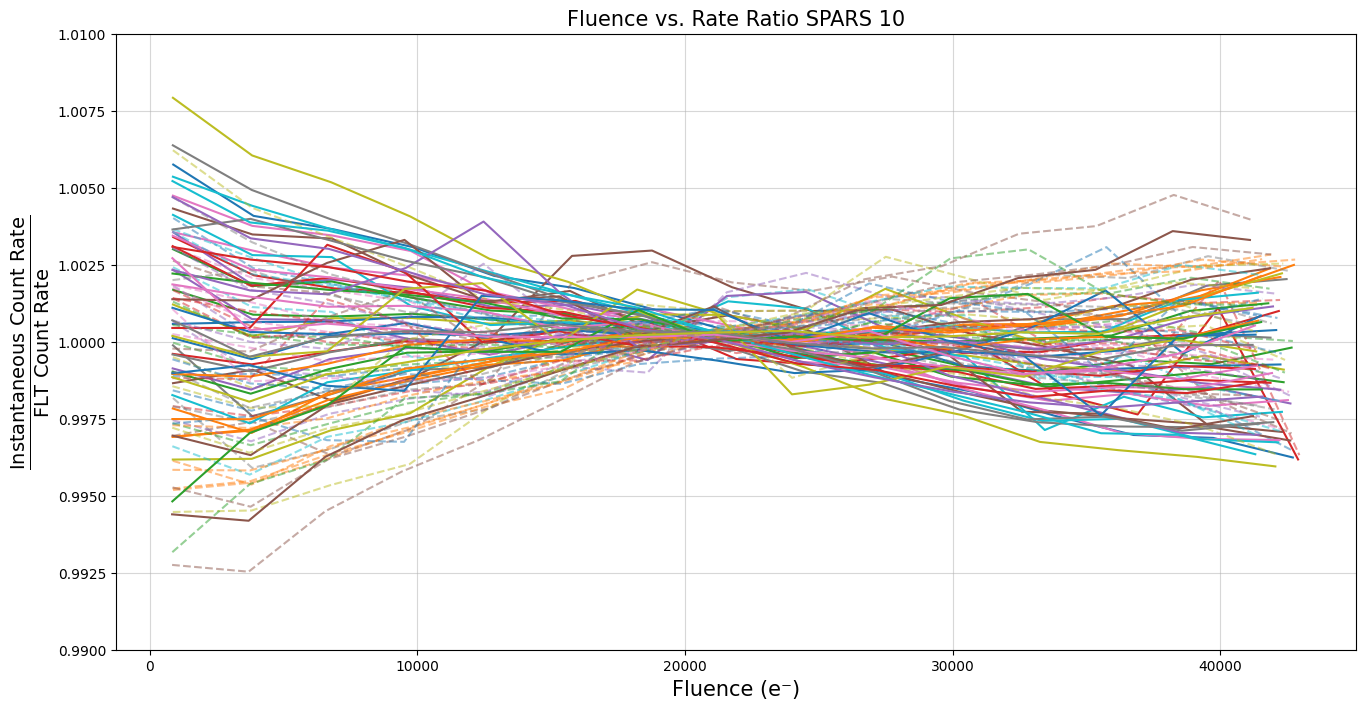}
\end{center}
\caption{\textit{Same as Figure \ref{fig:intflat_spars50}, but for a set of \texttt{SPARS10} INTFLATS that sample low fluence levels. The peak-to-peak range is \~1\% across all fluence levels for both the old and the new corrections.}
}
\label{fig:intflat_spars10}
\end{figure}

\begin{figure}[H]
\begin{center}
\includegraphics[width=.90\linewidth]{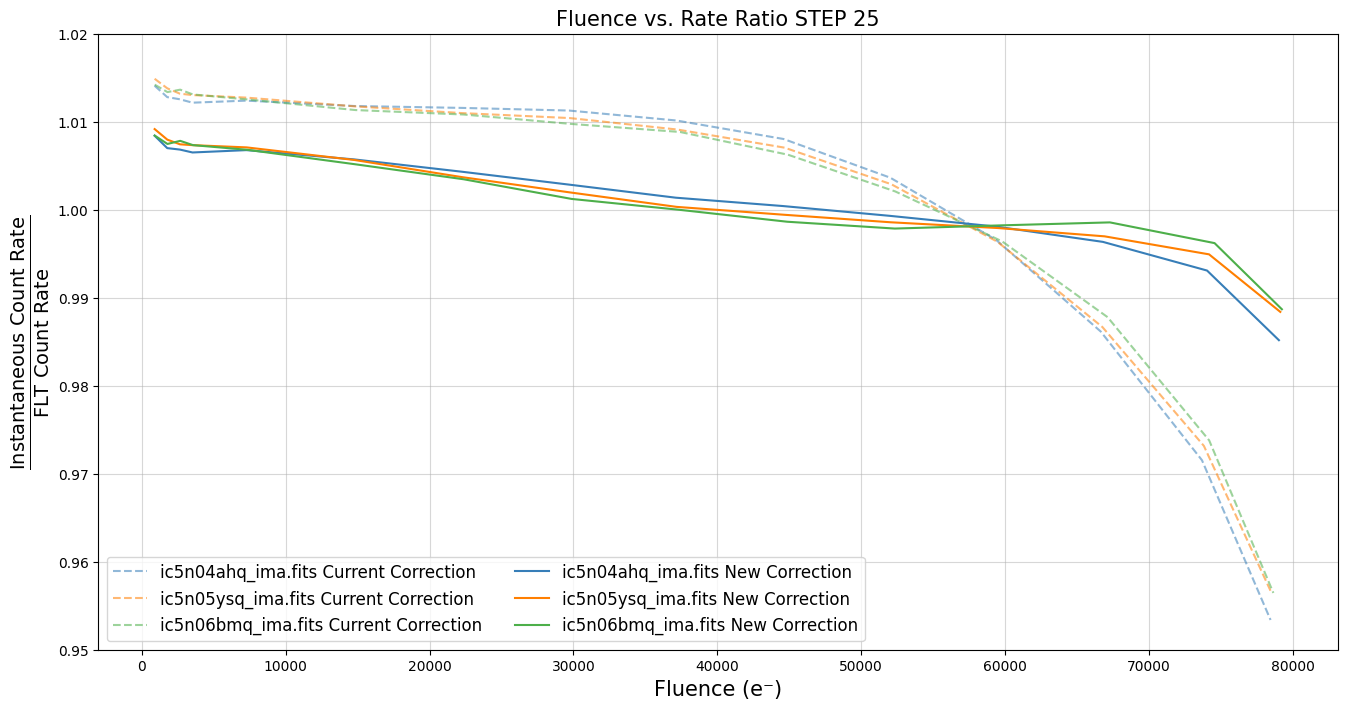}

\end{center}
\caption{\textit{Same as Figure \ref{fig:intflat_spars50}, but for a set of \texttt{STEP25} INTFLATS. The peak-to-peak range over all fluence levels is reduced from 5\% to 2\% when using the new linearity correction.}
}
\label{fig:intflat_step25}
\end{figure}

\begin{figure}[H]
\begin{center}
\includegraphics[width=.92\linewidth]{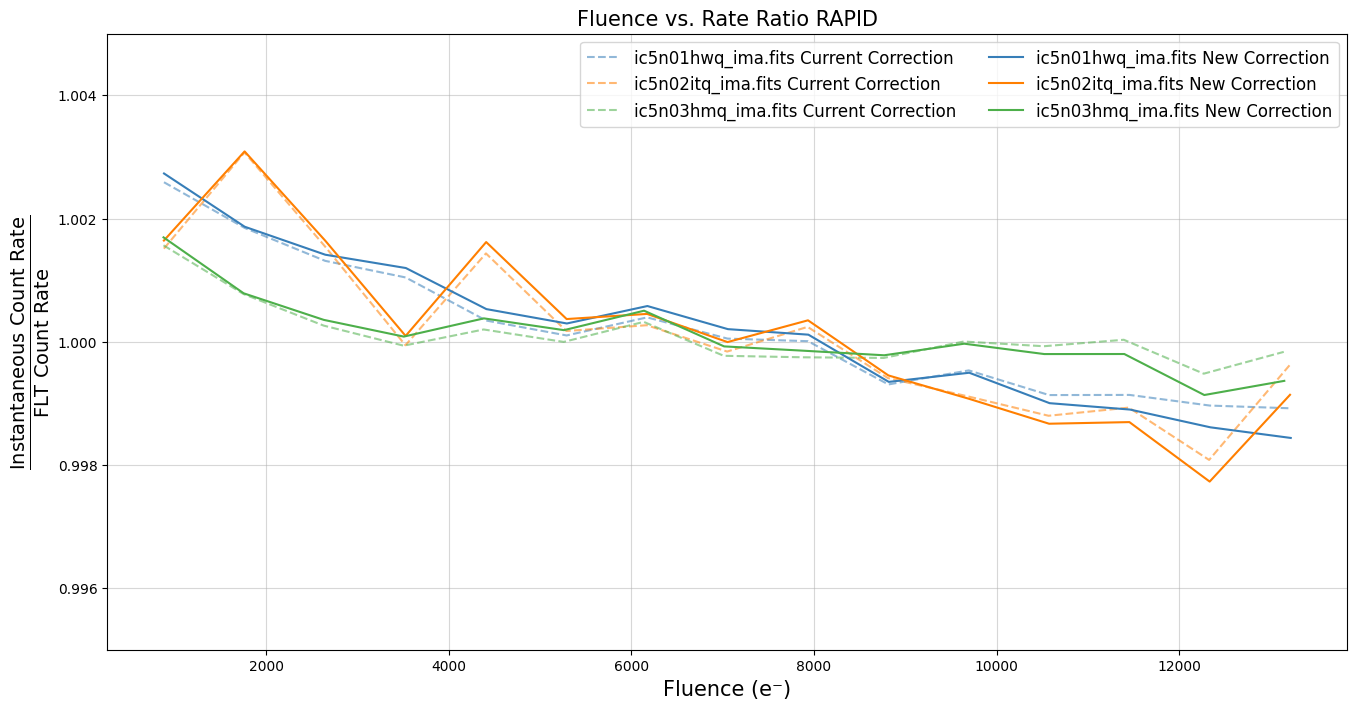}
\end{center}
\caption{\textit{Same as Figure \ref{fig:intflat_spars50}, but for a set of \texttt{RAPID} INTFLATS which sample the lowest fluence levels. The peak-to-peak range remains around 0.4\% for both the old and the new correction.}}
\label{fig:intflat_rapid}
\end{figure}

Internal lamp flats (INTFLATS) are useful for measuring detector non-linearity effects because they provide a stable, uniform source of illumination under repeatable conditions. This allows for a precise characterization of how the detector response deviates from a perfectly linear relationship at various fluence levels. We reprocess INTFLAT data taken as part of our annual non-linearity monitoring program with our new correction file in five different sample sequence modes, which sample a range of signal levels: \texttt{SPARS25} and \texttt{SPARS50} at high fluence levels, and \texttt{SPARS10}, \texttt{STEP25}, and \texttt{RAPID} at low fluence levels. Table \ref{tab:intflat} in the Appendix lists the sample sequences (\texttt{SAMP-SEQ}), the number of reads (\texttt{NSAMP}), the exposure time, rootnames, and proposal IDs of the INTFLAT exposures used to validate the new linearity correction.
\begin{figure}[H]
    \centering
    \includegraphics[width=.82\linewidth]{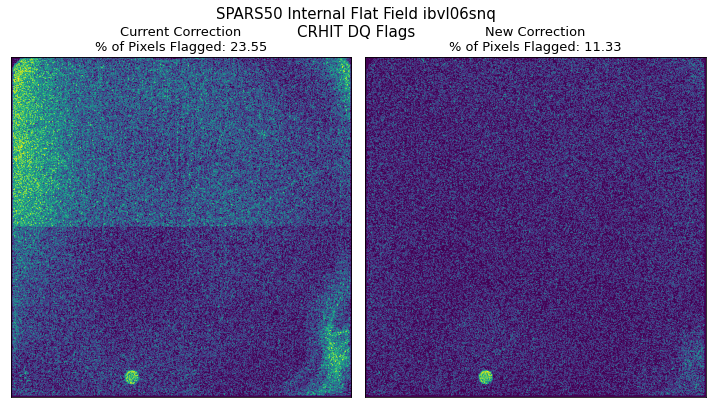}
    \includegraphics[width=.82\linewidth]{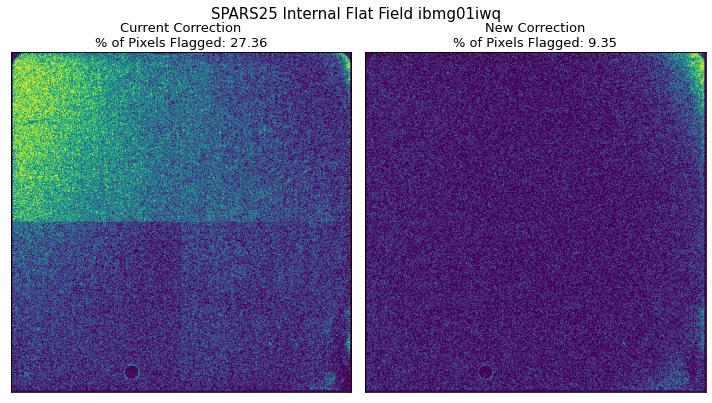}
    \caption{\textit{CRHIT DQ flags in the last \texttt{ima} read, e.g. \texttt{[DQ, 1]}, for internal flat fields (INTFLATS) acquired with different sample sequences:  \texttt{SPARS50} (\texttt{ibvl06snq}) and \texttt{SPARS25} (\texttt{ibmg01iwq}). For each INTFLAT, the left panel is calibrated with the 2009 quad-based \texttt{NLINFILE} and the right panel with the new pixel-based \texttt{NLINFILE}. Yellow points show the position of cosmic ray flags (DQ $\geq$ 8192) populated by calwf3 during the up-the-ramp fit. Each panel reports the percentage of detector pixels flagged, with the new solution flagging two times fewer pixels in the \texttt{SPARS50} INTFLAT, and three times fewer pixels in the \texttt{SPARS25} INTFLAT. With the new solution, the number of flagged pixels is now similar in all four quadrants. }}
    \label{fig:intflat_crflag}
\end{figure}
Cosmic ray flags (hereafter referred to as CRHIT, \texttt{DQ} = 8192) are flagged when the pixel's read value is found to experience a ``jump" (usually caused by a cosmic ray) and deviates from the linear fit to the calibrated ramp. These pixels are then rejected from the fitting process, with a rejection threshold of 4-$\sigma$  defined by the \texttt{CRREJTAB} reference table \parencite{DHB}. However, at levels close to saturation, the linear fit deviates from linearity, and thus may cause erroneous flagging of CRHITs in exposures.
\bigbreak
Figure \ref{fig:intflat_crflag} shows the CRHIT flags in the last \texttt{ima} read for a \texttt{SPARS50} and \texttt{SPARS25} INTFLAT calibrated with the new and current linearity corrections. For the \texttt{SPARS50} INTFLAT, the current linearity correction tends to overflag pixels, with $23\%$ of detector pixels flagged. This is reduced to $11\%$ of detector pixels with the new correction. For the \texttt{SPARS25} flat, the improvement is even larger, where the current correction flags nearly three times as many pixels ($27\%$) compared to the new correction ($9\%$). This discrepancy likely arises from limitations in the coefficients derived from the heavily saturated \texttt{STEP50} INTFLATs used to derive the current quadrant-averaged solution, as well as from unmodeled pixel-to-pixel variations within each quadrant. We also note that the majority of the flags are in quadrant 1 (the upper left quadrant of the IR channel). These flags highlight larger deviations in the true pixel-based linearity compared to the quadrant average. We note a spatial pattern in the number of DQ flags for quadrant 1, with the left third being flagged more frequently with the current correction. 
\bigbreak
We define the instantaneous count rate ($e^-/s$) between reads as the measured count rate accumulated between two consecutive reads: 
\begin{equation}
    \frac{Signal_N - Signal_{N-1}}{Time_N - Time_{N-1}}
\end{equation}
where \textit{N} is the read number, \textit{Signal} is the total measured signal of the read ($e^-$), and \textit{Time} is the time at which the read is sampled ($s$). We also define the fluence level ($e^-$) as the total accumulated signal at a given read:
\begin{equation}
Count\ Rate_N \ * Time_N
\end{equation}
Plotting the instantaneous count rate versus the fluence level provides a diagnostic of the IR channel’s linearity. For a perfectly linear detector, we expect the instantaneous count rate to be flat as the fluence level rises. This relationship between instantaneous count rate and fluence will be used throughout the remainder of this report to probe the improvements in our correction.
\bigbreak
We compute 1.) the ratio between the instantaneous count rate of each read in the \texttt{ima} files and the count rate derived from the \texttt{flt} files using the up-the-ramp fit, and 2.) the fluence level for the given ramp read across every pixel for all INTFLATs observed in the five different sample sequences. INTFLAT observations are measured in units of DN, so we take the additional step of multiplying the pixel values by the gain (2.5 $e^-/DN$) \parencite{2015gosmeyer}. We mask saturated pixels according to their DQ flag, then calculate the 3-$\sigma$ clipped mean of the count rate ratio and fluence across the entire flat for each read. Figures \ref{fig:intflat_spars50}--\ref{fig:intflat_rapid} show the clipped mean count rate ratio versus the clipped mean fluence level across every read for INTFLATs taken with \texttt{SPARS50}, \texttt{SPARS25}, \texttt{SPARS10}, \texttt{STEP25}, and \texttt{RAPID} sample sequences, respectively.
\bigbreak
For \texttt{SPARS50} and \texttt{SPARS25} INTFLATs, we exclude the last three and four reads, respectively, due to saturation. We find the most significant improvements in the peak-to-peak range (full range from the minimum to the maximum) across all fluence levels when using the new correction in \texttt{SPARS50}, \texttt{SPARS25}, and \texttt{STEP50} sample sequence modes, with a reduction from $3\%$ to $1\%$, $7\%$ to $2\%$ (the greatest improvement), and $5\%$ to $2\%$, for the three sample sequence modes, respectively. This reduction in peak-to-peak range indicates that the new correction improves both the bright and faint ends when compared to the current correction in these sample sequence modes. The peak-to-peak range remains the same between the two non-linearity correction files for both \texttt{RAPID} and \texttt{SPARS10}, varying by 1\% and 0.4\%, respectively.

\section*{Star Clusters}
\subsection*{NGC1851}
We reprocess deep exposures ($\sim900$ seconds) of a field about three arcmin southwest from the center of globular cluster NGC 1851 (distance $\sim12$ kpc), taken in F110W from GO proposal 16177 (PI: Calamida) with our new linearity correction file. 
Exposures in the visit are taken in non-interruptible sequence mode\footnote{Non-interruptible sequence (\texttt{NON-INT}) is a mode in HST's Astronomer's Proposal Tool (APT) which flags the observations to be executed back-to-back without any interruptions such as buffer dumps and gaps between Earth's occultation.} and dithered by 8-10 pixels via \texttt{POS-TARGS}, which limits persistence effects \parencite{2010Long}.

\begin{figure}[H]
    \centering
    \includegraphics[width=\linewidth]{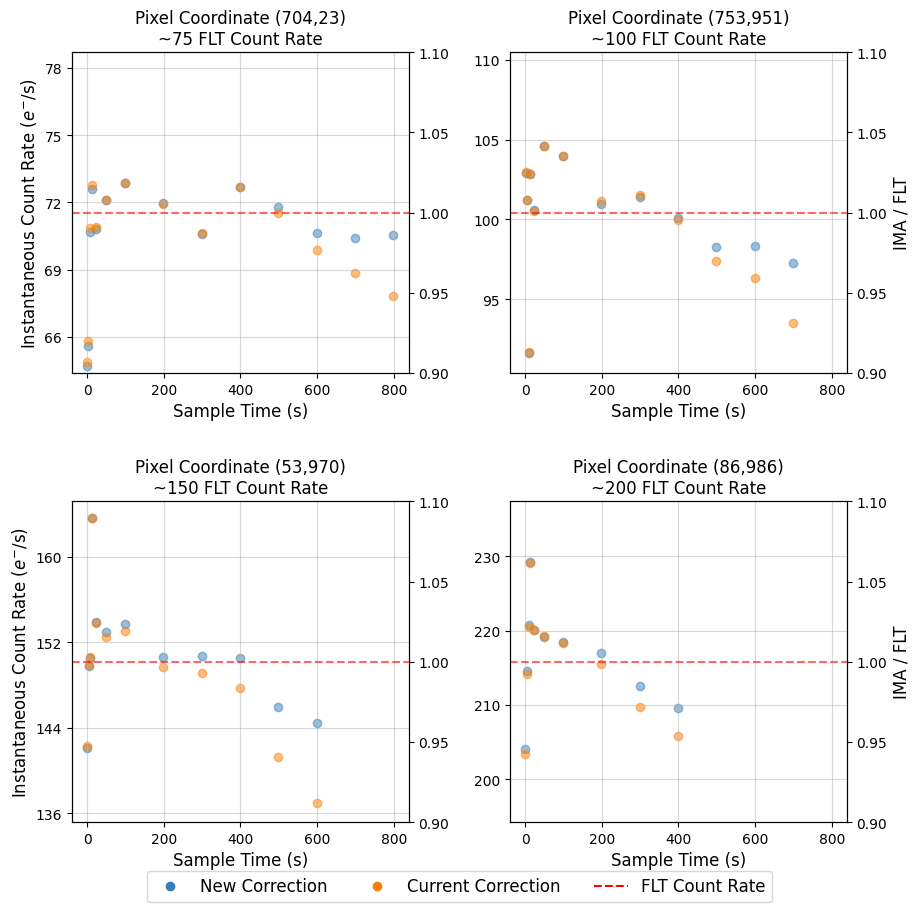}
    \caption{\textit{Instantaneous count rate ($e^-/s$) versus exposure time (s) for the peak pixel of four stars in the globular cluster NGC 1851, acquired in a $\sim$900-second exposure (\texttt{ieaa02e8q}) with the \texttt{STEP100} sample sequence. Each panel shows changes in the peak pixel value measured across different reads in the \texttt{ima} file for stars with different count rates: $\sim$75, 100, 150, and 200 $e^-/s$.  Orange points show the shape of the ramp calibrated with the 2009 linearity correction, while blue points show the improvement when using the new linearity correction. For reference, the \texttt{flt} count rate derived with the new correction is overplotted as a red dashed line. The 100, 150, and 200 count rate ramps reach saturation before the end of the exposure, so their final reads are not shown in the figure. In all four cases, the new linearity correction improves the results by $\sim1$--$4\%$ at the end of the ramp where the fluence approaches the full well limit.}}
    \label{fig:NGC1851_pixel_F110W}
\end{figure}

\begin{figure}[H]
\begin{center}
\includegraphics[width=\linewidth]{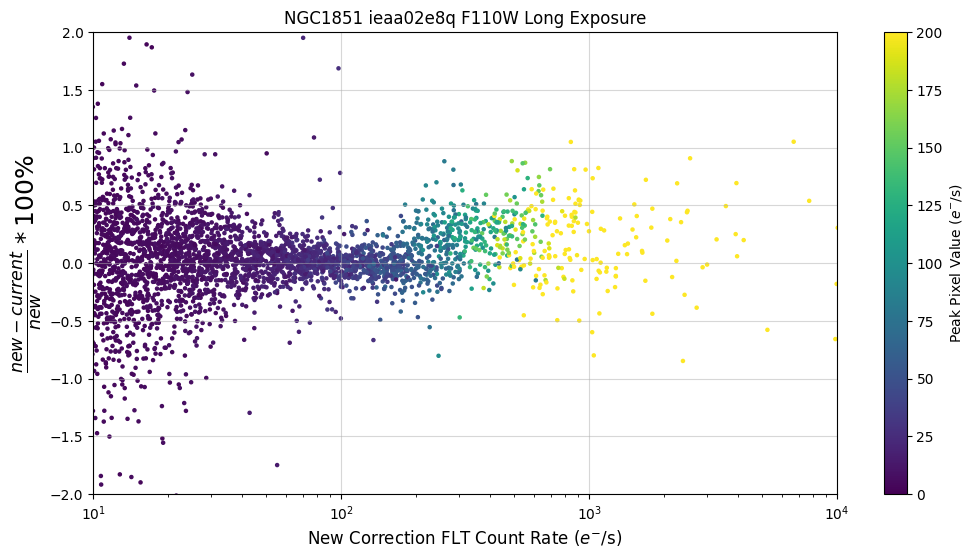}
\end{center}
\caption{\textit{Percent difference in the \texttt{flt} count rate ($e^-/s$) measured within a 3--pixel radius aperture for stars in NGC 1851 calibrated with the new and current (2009) \texttt{NLINFILE} plotted versus the \texttt{flt} count rate for the same exposure shown in Figure \ref{fig:NGC1851_pixel_F110W}. Points are colored based on their peak pixel value measured via the \texttt{hst1pass} software \parencite{Anderson2022}. At higher count rates, sources calibrated with the new non-linearity correction files are brighter by up to 0.5$\%$ in the 3-pixel aperture compared to the current correction.}}
\label{fig:NGC1851_F110W}
\end{figure}

\bigbreak
We identify sources with the \texttt{hst1pass} software routine \parencite{Anderson2022} for one F110W exposure \texttt{ieaa02e8q} (taken with \texttt{STEP100} sample sequence with 16 samples). Using the peak pixel value of sources found via \texttt{hst1pass}, we identify four sources at different count rates: $\sim$75, 100, 150, and 200 $e^-/s$. We then plot the instantaneous count rate of the \texttt{ima} ramps versus the instantaneous exposure time for the source with the current and new linearity correction file, as shown in Figure \ref{fig:NGC1851_pixel_F110W}. In all cases, at later reads where the pixel is approaching saturation, the new linearity correction improves the linearity by $\sim1$-$4\%$.
\bigbreak
We also perform 3--pixel radius aperture photometry on sources identified by \texttt{hst1pass} in the same exposure. We then plot the percent difference between the measured count rate in the aperture using the two correction files versus the measured count rate with the new correction as shown in Figure \ref{fig:NGC1851_F110W}. At higher count rates (100 $e^-/s$ and higher), the percent difference is 0--0.5\%, indicating that the current correction deviates from linearity at the bright end.
\bigbreak

\begin{figure}[H]
\begin{center}
\includegraphics[width=\linewidth]{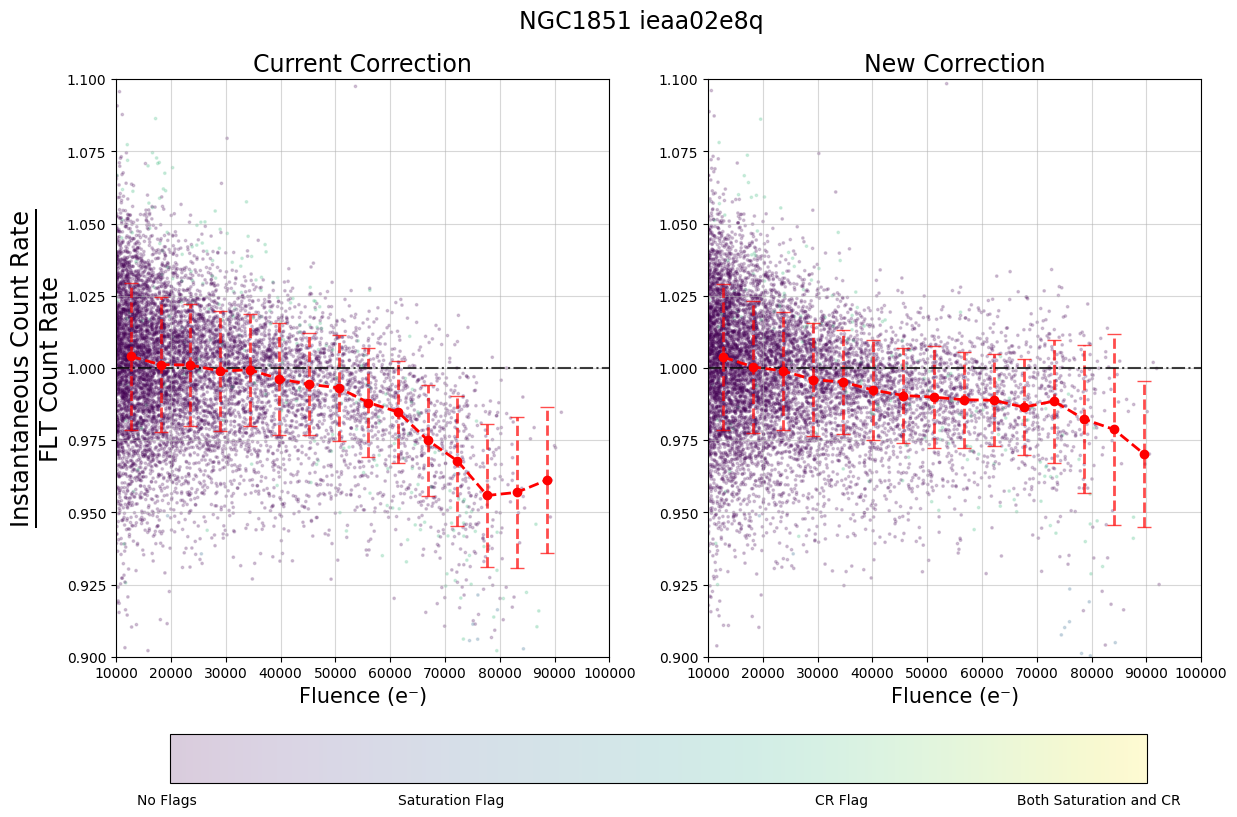}
\end{center}
\caption{\textit{Ratio of the instantaneous count rate of \texttt{ima} reads and the \texttt{flt} count rate versus the fluence level ($e^-$) for the peak pixel of sources identified by \texttt{hst1pass} for a $\sim$900-second exposure of NGC 1851 (\texttt{ieaa02e8q}) acquired in the F110W filter with the \texttt{STEP100} sample sequence. The two panels compare the results calibrated with 1.) the current 2009 non-linearity correction file (left) and 2.) the new pixel-based correction (right). Median binned statistics are overplotted for fluence levels between 10,000 and 100,000 $e^-$, where early reads are excluded due to their large scatter. Points are colored based on their DQ flag populated in the \texttt{ima} file during the ramp fitting step, where purple indicates no flag. Only about 1\% of sources measured were flagged with CRHIT in this exposure. The new correction improves the linearity by $\sim$1--2$\%$ compared to the current correction at higher fluence levels for sources identified by \texttt{hst1pass}.}}
\label{fig:ngc1851_instcr} 
\end{figure}

Figure \ref{fig:ngc1851_instcr} shows the ratio between the instantaneous count rate of \texttt{ima} reads and the \texttt{flt} count rate value versus the fluence level ($e^-$) for the peak pixel of sources identified by \texttt{hst1pass} with the current and new non-linearity correction (on the same exposure \texttt{ieaa02e8q}). We calculate and plot the median-binned statistic with \textit{n} bins = 15 for reads with a fluence level between 10,000--100,000 $e^-$, where early reads are excluded due to their scatter. The color of the points indicates whether a DQ flag was populated in the \texttt{ima} file during the ramp fitting step, where purple indicates no flag and green shows stars that were flagged as cosmic ray hits in their ramps. In the case of this exposure, only about 1\% of pixels are flagged as CRHIT. The current correction deviates from perfect linearity (a count rate ratio of $\sim$=1) at higher fluence levels by $\sim4\%$, while the new correction improves the deviation from linearity by $\sim$1--2$\%$, reinforcing the improvement at higher fluence levels when using the new pixel-based correction.

\subsection*{47 Tucanae}
\begin{figure}[H]
    \centering
    \includegraphics[width=\linewidth]{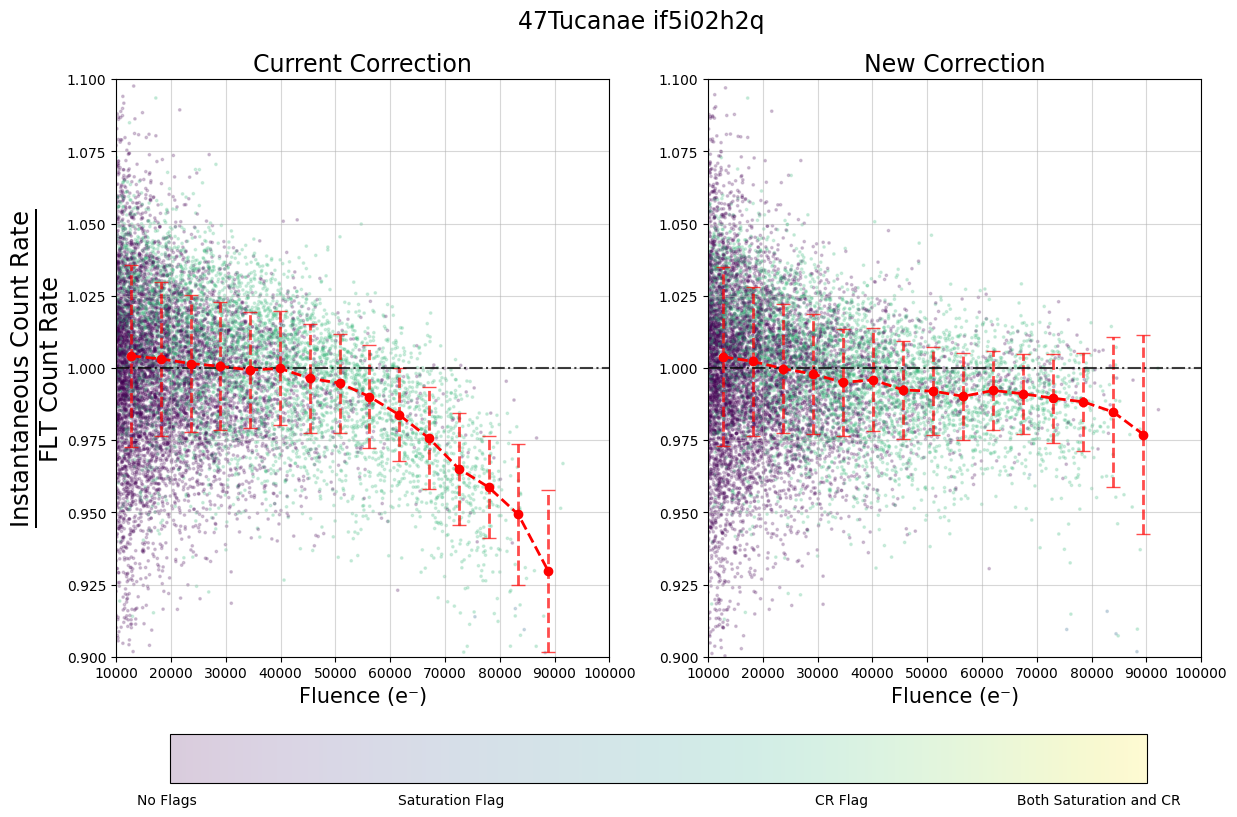}
    \caption{\textit{Same as Figure \ref{fig:ngc1851_instcr}, but for a $\sim$350-second exposure of 47 Tucanae (\texttt{if5i02h2q}) acquired in the F160W filter with the \texttt{SPARS25} sample sequence. The current (2009) correction deviates from linearity at higher fluence levels by up to $\sim7\%$, while the new correction deviates only by $\sim2\%$ at the same fluence levels, significantly improving the linearity of the peak pixel of sources. The colorbar indicates whether a flag was populated in the DQ array of the \texttt{ima} file, where purple indicates no flag and green highlights pixels flagged as a cosmic ray by \texttt{calwf3} during the ``up-the-ramp" fit. [\textsc{Note added in proof}: Red points in the right panel were updated with the publication of the Addendum, after discovery of a binning error in the originally published ISR.]
    }}
    \label{fig:47tuc_instcr}
\end{figure}

We repeat the analysis done with NGC 1851 in Figure \ref{fig:ngc1851_instcr} with 47 Tucanae (a globular cluster at $\sim$4.5 kpc) using \texttt{if5i02h2q}, a $\sim$350-second F160W exposure acquired with the \texttt{SPARS25} sample sequence from the WFC3 calibration (CAL) program 17363. While the goal of this program is to monitor the WFC3/IR time-dependent sensitivity with cluster observations \parencite{2022Bajaj}, this program duplicates the observing strategy from the prior ``WFC3/IR Signal Non-Linearity Monitor," which was designed to test the linearity correction derived from internal flats using star cluster data (see programs 12352, 12696, 13079, and 13563) and thus is useful for our purpose here.
\bigbreak
Figure \ref{fig:47tuc_instcr} shows the ratio of the instantaneous count rate of \texttt{ima} reads and the \texttt{flt} count rate versus the fluence level ($e^-$) for the peak pixel of sources identified by \texttt{hst1pass} for 47 Tucanae exposure \texttt{if5i02h2q}. The two panels show the results calibrated with the current and new non-linearity corrections (same analysis as Figure \ref{fig:ngc1851_instcr}). Points are colored based on their DQ flag in the same way as Figure \ref{fig:ngc1851_instcr}. The current correction deviates from linearity at higher fluence levels by up to $\sim7\%$, while the new correction significantly improves the linearity, deviating only by $\sim2\%$ at saturated fluence levels. 
We noticed a large number of pixels flagged as a CRHIT in this exposure, which was observed using \texttt{SPARS25} sample sequence, the same sample sequence shown in Figure \ref{fig:intflat_crflag}. We repeat the analysis in Figure \ref{fig:47tuc_instcr} for stars in Quadrants (Quads) 1 and 4, which comprise the top half of the IR detector. We find that the quad-based non-linearity correction is significantly worse in Quad 1 with deviations up to $\sim8\%$, compared to Quad 4 with deviations up to $\sim5\%$. This strongly implies that the larger number of CR flags in Quad 1 is related to errors in the quadrant-based linearity solution. 
\bigbreak
This exposure also has $\sim$15\% more CRHIT flags identified by \texttt{calwf3} than that of the NGC 1851 exposure in Figure \ref{fig:ngc1851_instcr}. This is most likely related to the different sample sequence of the two exposures. The NGC 1851 exposure in the previous section was observed in \texttt{STEP100} (linearly spaced readouts early in the ramps ending with logarithmically spaced readouts) while the 47 Tucanae exposure is observed in \texttt{SPARS25} (evenly spaced readouts). The smaller spacing between 47 Tucanae reads may cause \texttt{calwf3} to erroneously detect pixels as a CRHIT more often compared to a larger time step between reads. We compare the total number of CR flagged peak pixels of sources in Quadrants 1 and 4 in the 47 Tucanae exposures with the two corrections and find that there is a 20\% reduction in the number of peak pixels being flagged as CRHIT when using the new correction compared to the current one, while the number of CRHIT flags in Quadrant 4 did not change significantly.

\section*{Staring Mode Observations of  Standard Star P330E}
\begin{figure}[H]
    \centering
    \includegraphics[width=.84\linewidth]{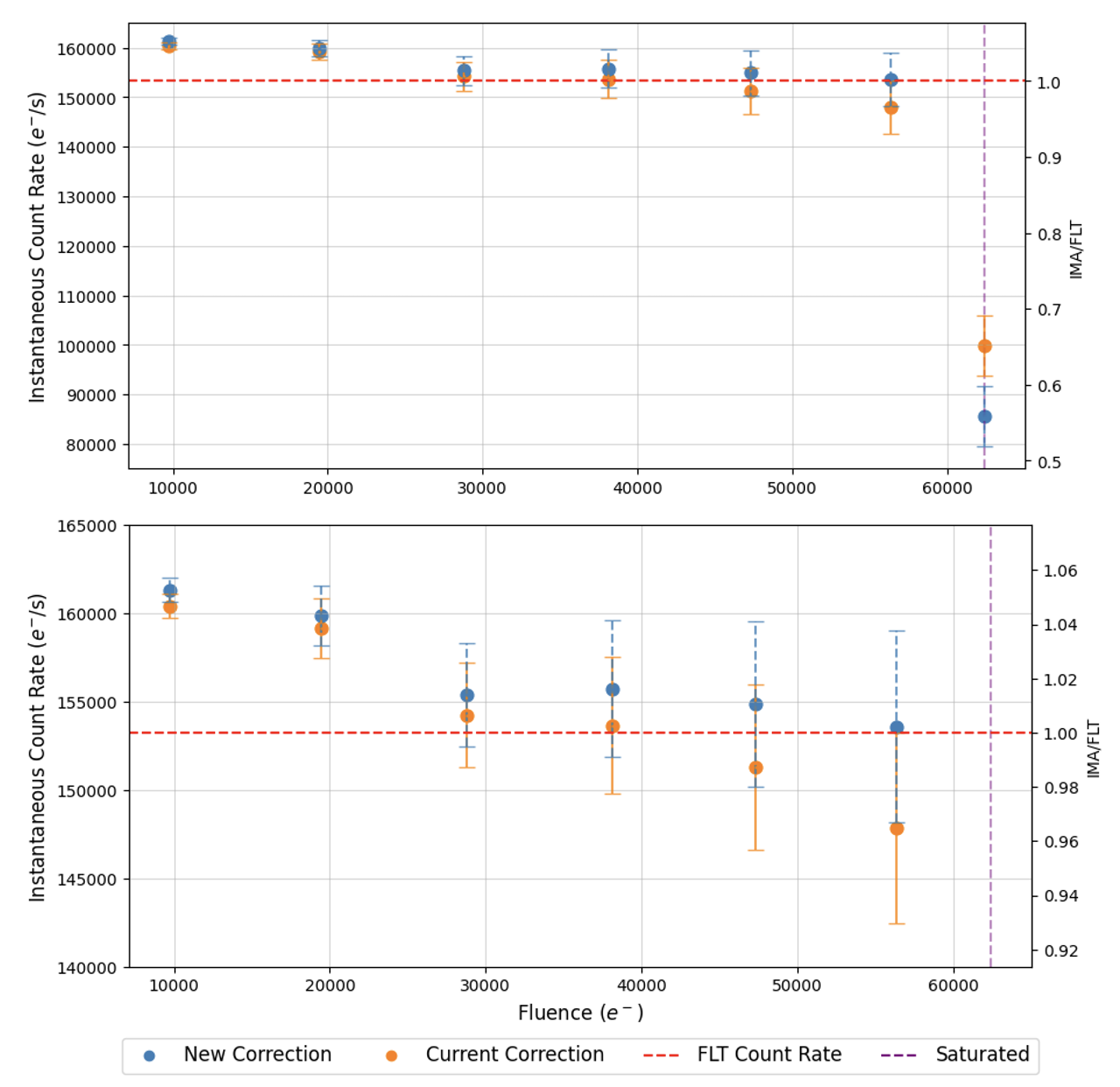}
    \caption{\textit{Instantaneous count rate ($e^-/s$) versus the fluence level ($e^-$) for the peak pixel value in  F110W P330E exposure \texttt{ie7h20bvq} corrected with the new non-linearity correction file (blue) and the current (2009) correction file in the pipeline (orange). The count rate measured in the \texttt{flt} file (red dotted line) and the fluence level at which saturation is first flagged in the DQ array (purple dotted line) is also shown. The error bars represent the propagated uncertainties of the instantaneous count rates derived from the \texttt{ERR} extensions of the \texttt{ima} reads. At ramps approaching saturation, the new correction improves the linearity compared to the current correction, most notably a $\sim3\%$ improvement in the last ramp before saturation begins at a fluence level of $\sim55,000$ $e^-$. The bottom panel is identical to the top except for a smaller y-axis range.}}
    \label{fig:P330E_single}
\end{figure}

High signal-to-noise observations of spectrophotometric standard stars serve as an excellent monitor of the IR channels' throughput and stability. P330E is a G-type CALSPEC standard star used for flux calibration monitoring due to its well-characterized spectral energy distribution. We reprocess P330E observations across three filters (F110W, F140W, and F160W) from past WFC3 photometric monitoring programs spanning from cycles 17 - 30 (2009 - 2022) to test our new linearity correction file. All observations are taken with \texttt{RAPID} sample sequence with a mix of small subarray apertures to avoid saturation, while still providing a large number of reads to obtain an accurate ramp fit. Ramps with four or fewer reads are removed from our analysis. Table \ref{tab:p330etab} in the Appendix lists the Proposal ID, filter, sub-array, aperture used, exposure time (s), and rootnames of the P330E exposures.
\bigbreak
Figure \ref{fig:P330E_single} shows the instantaneous count rate ($e^-/s$) versus the fluence level ($e^-$) for the peak pixel value as determined by \texttt{hst1pass} of one P330E exposure \texttt{ie7h20bvq} in F110W corrected with the current and new correction file. This exposure was acquired with the \texttt{RAPID} sample sequence using the \texttt{IRSUB64} subarray in order to acquire a large number of reads (\texttt{NSAMP} = 8) in the short 0.4-second exposure before the star saturates. In the image, the star was positioned at the center of a pixel and therefore reaches the saturation limit of the pixel, whereas other images of this star do not get close to the full well limit due to the star being centered at the corner of four pixels and the flux of the peak pixel is spread out. The error bars represent the propagated uncertainties of the instantaneous count rates derived from the \texttt{ERR} extensions of the \texttt{ima} reads. At later \texttt{ima} reads, where the fluence level is close to saturation, the new correction (blue) brings the count rates of the higher ramps closer to linearity with the earlier ramps compared to the current correction (orange). Most notable is a $\sim3\%$ improvement in the last read before saturation begins compared to the measured \texttt{flt} count rate, validating the improvements with the new pixel-based solution.

\begin{figure}[H]
\begin{center}
\includegraphics[width=.92\linewidth]{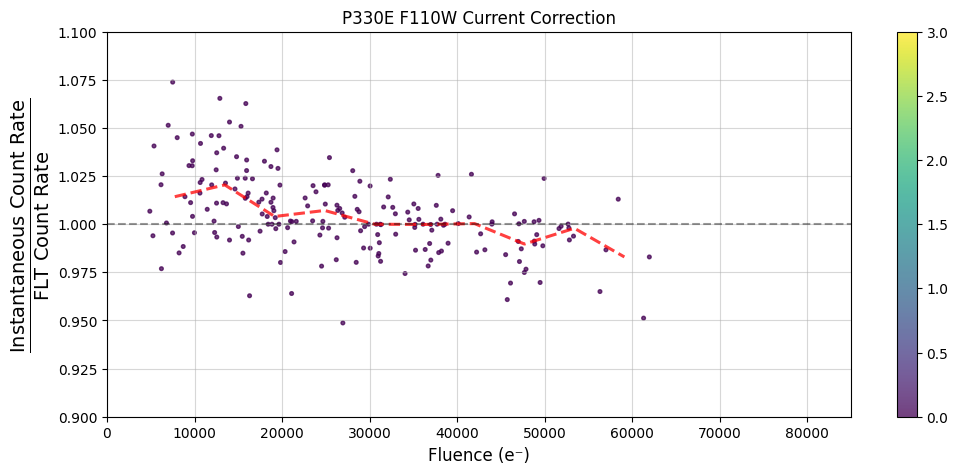}
\includegraphics[width=.92\linewidth]{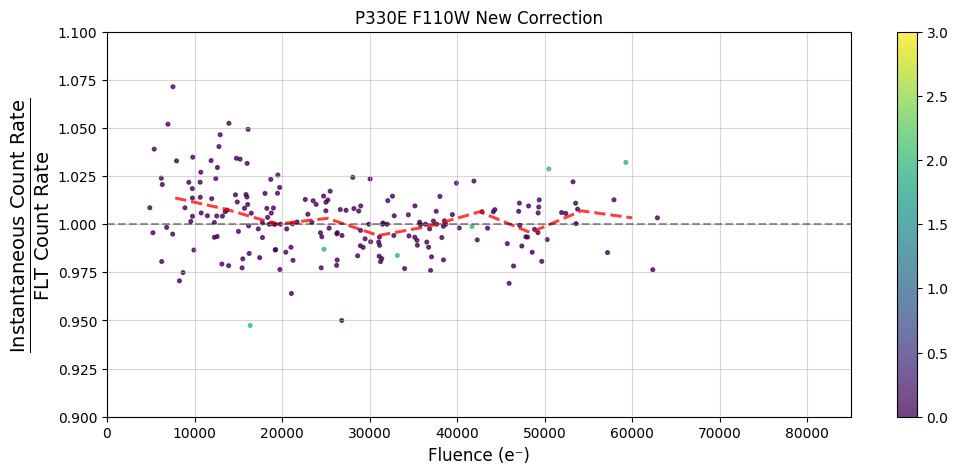}
\end{center}
\caption{\textit{Ratio of the instantaneous count rate of \texttt{ima} reads and the \texttt{flt} count rate versus the fluence level ($e^-$) for P330E observations taken in F110W with the current (2009) linearity correction (top) and the new linearity correction (bottom). The colorbar presents the DQ flags of the pixel: purple (value = 0) = no flags, dark blue (1) = saturated pixel, light green (2) = CR flag, and yellow (3) = both saturated and CR flag. The median-binned statistic (red dashed line) is overplotted, and a horizontal line at a count rate ratio of 1.0 represents near-perfect linearity. The new correction file improves the linearity of P330E count rate measurements by $\sim2\%$ at higher fluence levels in F110W.}}
\label{fig:p330e_F110}

\end{figure}

\begin{figure}[H]
\begin{center}
\includegraphics[width=\linewidth]{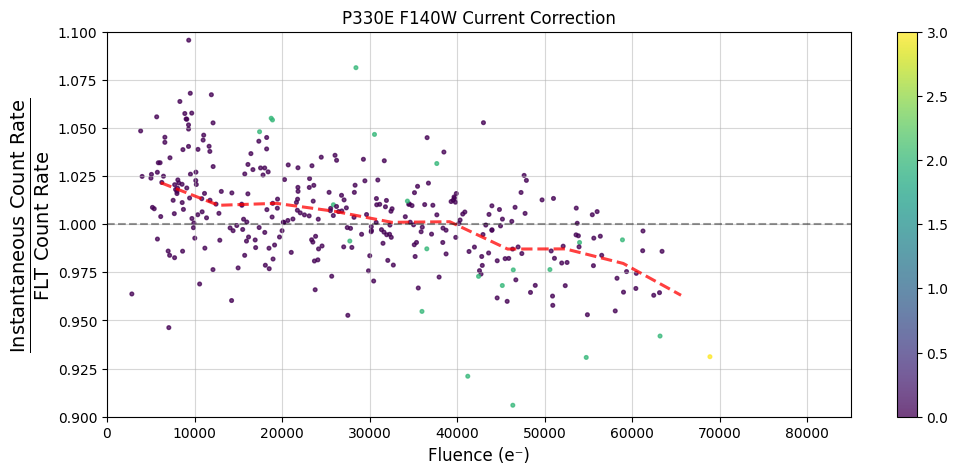}
\includegraphics[width=\linewidth]{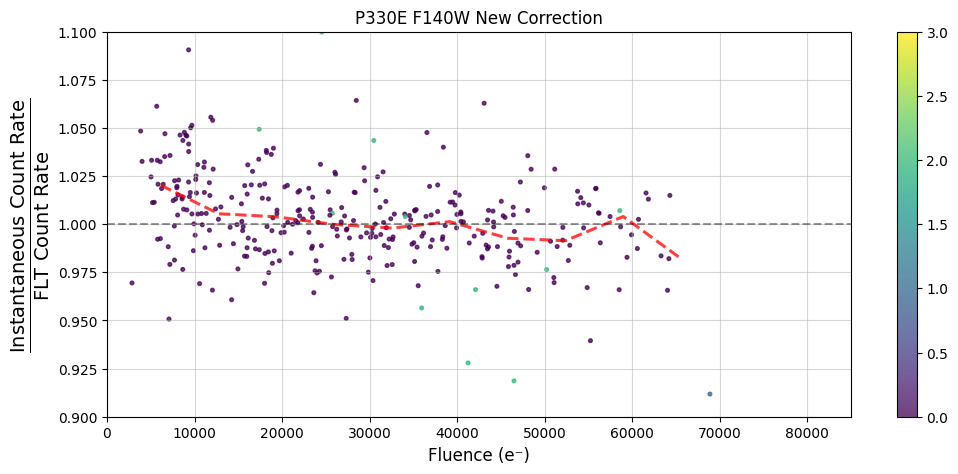}
\end{center}
\caption{\textit{Same as Figure \ref{fig:p330e_F110}, but for F140W. The new correction file improves the linearity of P330E count rate measurements by $\sim2\%$ at higher fluence levels in F140W.}}
\label{fig:p330e_F140}

\end{figure}

\begin{figure}[H]
\begin{center}
\includegraphics[width=\linewidth]{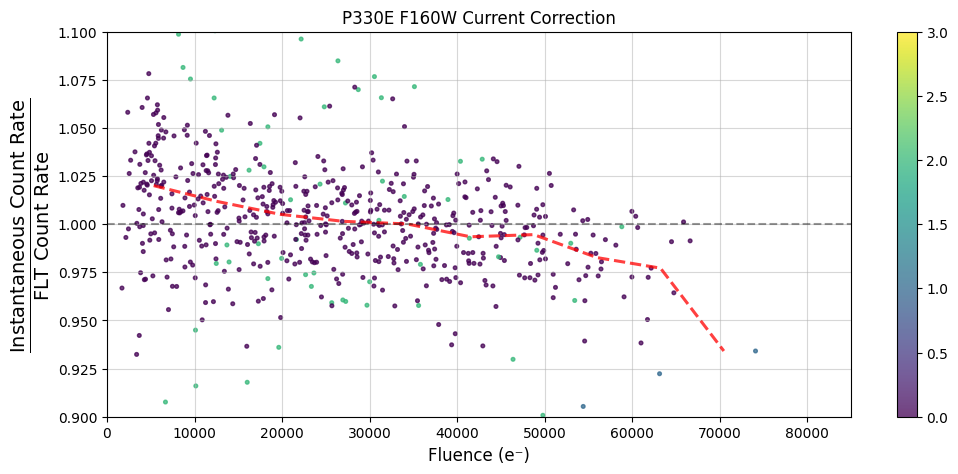}
\includegraphics[width=\linewidth]{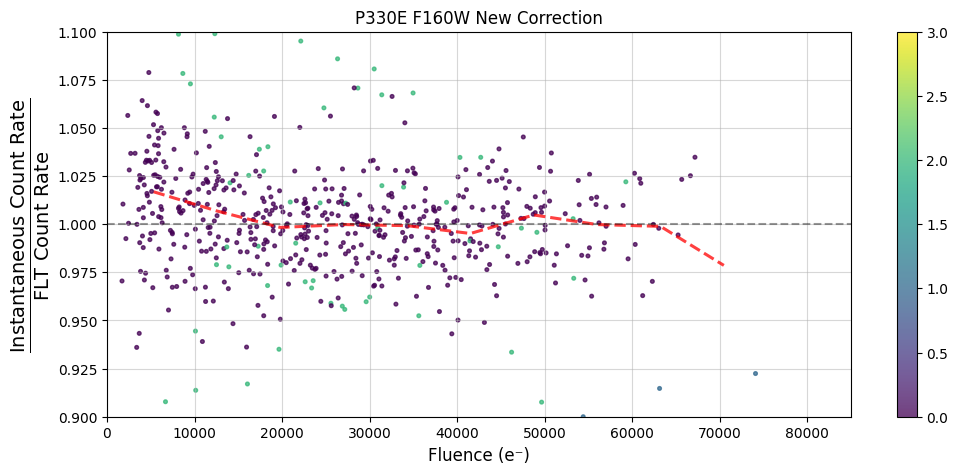}
\end{center}
\caption{\textit{Same as Figure \ref{fig:p330e_F110}, but for F160W. The new correction file improves the linearity of P330E count rate measurements by $\sim4\%$ at higher fluence levels in F160W.}}
\label{fig:p330e_F160}

\end{figure}

Furthermore, we repeat a similar analysis to that conducted on the INTFLAT observations in the earlier section. We measure the ratio between the instantaneous count rate between \texttt{ima} reads and the \texttt{flt} count rate, as well as the fluence level at the peak pixel coordinate identified by \texttt{hst1pass} of P330E only (as opposed to measuring every pixel observed and then calculating the sigma-clipped mean, as we did with the INTFLATs). Figures \ref{fig:p330e_F110}, \ref{fig:p330e_F140}, and \ref{fig:p330e_F160} show the count rate ratio versus the fluence level for each read of every P330E ramp for F110W, F140W, and F160W, respectively. The median-binned statistic with n bins = 10 is also plotted. Points are colored based on their DQ flags in the \texttt{ima} extension, where purple = no flags, dark blue = saturated pixel, light green = CR flag, and yellow = both saturated and CR flag. 

In all three filters tested, the new correction file improves the linearity, especially at the bright end. At higher fluence levels, the new correction shows the count rate ratio approaches 1.0—indicating near-perfect linearity—by 2--4\% more than the current correction across all three filters, with the most significant improvement of 4\% in F160W.
\section*{Grism Observations of CALSPEC Standard Stars}
\subsection*{P330E}
\begin{figure}[H]
    \centering
    \includegraphics[width=\linewidth]{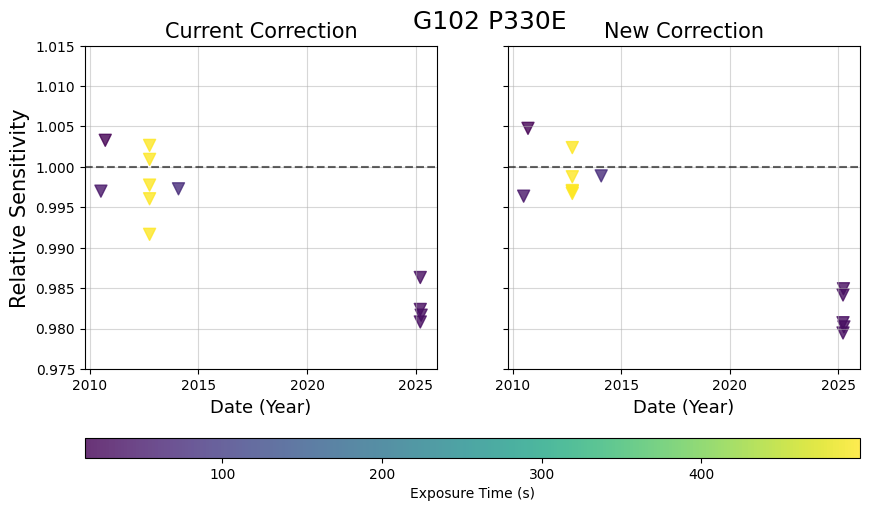}
    \caption{\textit{Relative sensitivity versus date for G102 observations of P330E calibrated with the current (2009; left) and new (right) non-linearity correction. Points are colored based on their exposure time. There is an improvement in the scatter for yellow points in $\sim$2012, where the long 500-second exposures taken with the \texttt{STEP50} sample sequence are closer to full well saturation than other visits. Purple points denote observations in a variety of sample sequences and exposure times ranging from 14 to 73 seconds.}}
    \label{fig:g102p330e}
\end{figure}

\begin{figure}[H]
    \centering
    \includegraphics[width=\linewidth]{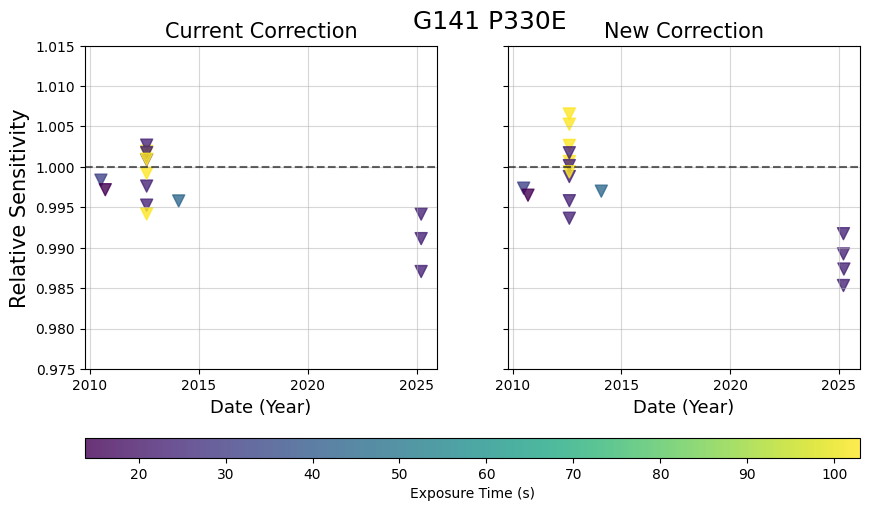}
    \caption{\textit{Same as Figure \ref{fig:g102p330e}, but for G141. While there is no significant improvement in the scatter for a given visit, the new correction increases the brightness of the yellow points (103 seconds, \texttt{SPARS10}) by $\sim0.5\%$. These data are affected by self-persistence from prior exposures in purple (23 sec, \texttt{RAPID}) due to a lack of dithering between pairs of long and short observations.}}
    \label{fig:g141p330e}
\end{figure}

IR grism observations of P330E (G102 and G141) were reprocessed with the current and new non-linearity corrections (Bohlin 2025, private communication) using the methodology described in \textcite{2019Bohlin}. These grism observations are part of the WFC3/CAL IR Grism Flux Monitor and were recently used to measure the time-dependent sensitivity of the IR detector \parencite{Som2024}. Table \ref{tab:p330egrism} in the Appendix lists the proposal ID, rootname, grism filter, and exposure time for the observations used.
\bigbreak
Figures \ref{fig:g102p330e} and \ref{fig:g141p330e} show the relative sensitivity over time based on P330E observations in G102 and G141 with the current and new non-linearity correction, respectively. The points are color coded by their exposure time. In G102, we see an improvement in the scatter ($\sim$0.001 improvement in the RMS when using the new correction) among observations during the same visit in the 2012 series, where the exposure (500 sec) was closer to saturation than in other exposures. This is consistent with previous tests showing that the correction improves the linearity of the count rate close to the saturation limit.

 However, for G141, the correction does not improve the linearity of the count rates for the 2012 visit, and the results actually have larger scatter. This visit in particular is affected by self-induced persistence to a greater degree than other visits, and its results may be inconclusive. However, the change between the two corrections in all G141 observations (including the one affected by persistence) is negligible ($<0.1\%$).

\subsection*{GD71}
\begin{figure}[H]
    \centering
    \includegraphics[width=\linewidth]{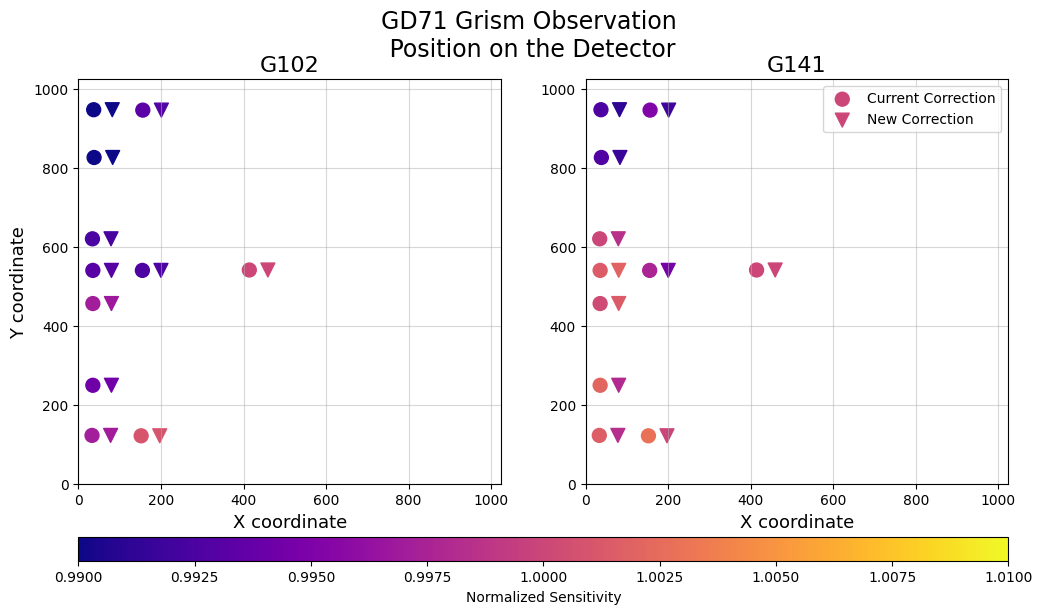}
    \caption{\textit{(X,Y) coordinate of GD71 taken in with \texttt{SPARS25} sample sequence for dithered exposures of G102 and G141 calibrated with the current (2009; circle) and new (triangle) non-linearity correction files. Points are colored based on their sensitivity, normalized to an image at the IR grism reference position near the center of the detector. Results from data processed with the new correction are shifted by 30 pixels along the X-axis to help compare differences with the previous correction. In general, the count rates are not significantly different between the old and new linearity corrections. The upper left corner of Quad 1 has lower observed count rates compared to the center of the detector, even with the new linearity correction.}}
    \label{fig:gd71pos}
\end{figure}

To test the spatial dependence of the linearity calibration, dithered observations of the CALSPEC standard star GD71 acquired in the G102 and G141 grisms were reprocessed with the new linearity solution (Bohlin 2025, private communication). These observations are part of the IR grism flux monitor calibration program and have recently been used to measure sensitivity losses of the IR detector as a function of wavelength \parencite{Som2024}. For observations acquired on the same date (but on different regions of the detector), the large (1-2\%) scatter in photometry was originally attributed to errors in the flat field. Table \ref{tab:gd71pos} in the Appendix lists the rootname for the set of dithered GD71 observations acquired in 2015, with a single visit for each IR grism. 
\bigbreak
 Figure \ref{fig:gd71pos} plots the (X,Y) position of the dithered exposures taken in each visit. Observations calibrated with the current (circle) and new (triangle) linearity corrections are colored by their normalized sensitivity to the value at the reference position near the center of the detector. Results from data processed with the new correction are offset by 30 pixels along the X-axis to help compare the difference between the two non-linearity corrections.
\bigbreak
The sensitivity of GD71 in both filters, when calibrated with both correction reference files, is nearly identical at all parts of the detector measured. We would expect the sensitivity to improve on the left edge of the detector with the new correction, as the current correction has an abnormality where it flags pixels as CRHIT more often due to the saturated flats used in the derivation of the current correction (see Figure \ref{fig:intflat_crflag}). However, we see a negligible change in sensitivity, indicating that the left edge of the detector may still experience the same non-linearity issues with the new pixel-based correction. In addition, it is important to note that there is no evidence of the non-linearity becoming worse with the new correction.
\section*{IR Zeropoints}

In late 2024, the first set of time-dependent WFC3/IR inverse sensitivities (zeropoints) was delivered to the HST Calibration Reference Data System (CRDS) \parencite{Calamida2024} and installed in the automated calibration pipeline. 
These zeropoints are important for users to obtain accurate photometry, as WFC3/IR has experienced a 1--2\% total decline in photometric sensitivity (depending on filter) since installation in 2009 \parencite{2024Marinelli}. 
In this section, we present relative photometry from \texttt{flt} files of staring mode observations spanning July 2009 to August 2025 for the CALSPEC standards used in \textcite{Calamida2024}.\footnote{G191B2B, GD-153, GD-71, GRW70, and P330E.}
We address how the updated linearity solution affects this photometry to determine if the IR zeropoints should be recalculated and redelivered. 

\input{obsobs_p330e}

To test the impact of the linearity solution, we recalibrated all observations of the CALSPEC standards using both the new and current linearity solutions. 
In Figure \ref{fig:obsobs_p330e}, we show the effects of the linearity solution on  3-pixel aperture photometry for P330E in the F110W filter. 
We take the observed count rate from exposures calibrated with the new linearity solution ($F_{new}$) and divide by the observed count rate of the data calibrated with the current linearity file ($F_{current}$). 
The resulting ratio ($F_{new} / F_{current}$) is plotted as a function of observation date, and color-coded according to the observation exposure time. 
A solid line is plotted at $1.0$, indicating no change in the count rates between the new and current linearity solutions. 
Generally, photometry on files calibrated with the new linearity solution yields higher measured count rates than photometry of files using the current calibration. 
We overplot two statistics to illustrate the behavior of this subset: the $3\sigma$-clipped mean ($\,\overline{x}_{3\sigma}\,$) and the weighted mean ($\,\overline{x}_{w}\,$; weighted by photometric error). 
For this target/filter combination, the $3\sigma$-clipped mean and weighted agree within $0.03\%$, and are both $\approx 0.35 \%$ above unity. 
\input{obsobs_gd153}

Figure \ref{fig:obsobs_gd153} also plots F110W flux ratios as a function of observation date, but for GD-153. 
While calibrating the staring mode exposures with the new linearity solution results in slightly higher photometric counts than the current calibration, the increase in observed count rate for this star is less than that observed for P330E in Figure \ref{fig:obsobs_p330e}.
We find the $3\sigma$-clipped mean to be $\approx 0.1 \%$ and the weighted mean at $\approx 0.2 \%$, underscoring an important detail. 
Sigma-clipping a sample removes outliers, and in this case, produces a more robust estimation of the mean flux ratio within the distribution of flux ratios. 
Because $\overline{x}_{w}$ is weighted by the photometric errors, it better captures the precision of the underlying data. 
\bigbreak
Following the same process as in Figures \ref{fig:obsobs_p330e} and \ref{fig:obsobs_gd153}, we calculate the flux ratio $F_{new} / F_{current}$ for all observations. 
Filter/target combinations with a sample size less than five were excluded from further analysis.
In the top subplot of Figure \ref{fig:obs_percentdiff}, we plot the target-and-filter-specific weighted mean as a function of filter pivot wavelength. 
G191B2B mean ratios are plotted as yellow downward-triangles, GD-153 as red diamonds, GD71 as blue squares, GRW70 as black circles, and P330E as green right-facing triangles.
Therefore, the pink dotted line from Figure \ref{fig:obsobs_p330e} corresponds to the green triangle at $(1153.4, 1.0033)$, and the same line in Figure \ref{fig:obsobs_gd153} matches the red diamond at $(1153.4, 1.0021)$. 
A line is drawn at $1.0$, which, as in Figures \ref{fig:obsobs_p330e} and \ref{fig:obsobs_gd153}, would indicate perfect agreement between the photometry using the new linearity solution and photometry using the current calibration. 
Notably, all points are above this line: for all target/filter combinations, using the new linearity file results in higher count rates than the current linearity solution. 
\bigbreak
Within each filter, we then pooled the flux ratios for all stars and calculated the weighted mean (weighted by photometric errors); results are shown in the bottom subplot of Figure  \ref{fig:obs_percentdiff}. 
The filter-dependent weighted means are plotted as a function of pivot wavelength, and are represented by teal circles. 
Asymmetric error bars on those points indicate the median absolute deviation as calculated for values above and below the weighted mean. 
As in Figures \ref{fig:obsobs_p330e} and \ref{fig:obsobs_gd153}, we plot a line at $1.0$, and draw the overall weighted mean of all targets and stars ($\overline{x}_{w}$) as a horizontal pink dotted line. 

\input{obsobs}

Overall, calibrating with the new linearity solution increases the observed count rate. 
On a per-star, per-filter basis, as shown in the top subplot of Figure \ref{fig:obs_percentdiff}, the increase ranges from $\sim 0.05$--0.5$\%$, and can be sensitive to sample size effects. 
For example, there are only nine F110W exposures of G191B2B, yielding a weighted mean flux ratio of $\sim 0.5\%$. 
When data from all targets are pooled together, the weighted means of the flux ratios (teal points in the bottom subplot of Figure \ref{fig:obs_percentdiff}) appear somewhat correlated with filter: bluer filters demonstrate a $0.2-0.3\%$ change, but in the reddest filters and all narrow-band filters, the offset is $< 0.1 \%$. 
These data may be influenced by underlying factors such as the shape of the point spread function (PSF) and the width of the filter bandpass.
\bigbreak
PSFs in bluer filters are sharper, making the central pixel more likely to hit fluences in the regime ($>50,000 e^{-}$) for which the current linearity solution performs most poorly. 
In that case, the new linearity solution would have a more significant effect on the photometry, thereby driving the mean flux ratio up.
\bigbreak

By definition, narrow-band filters transmit only a narrow band of wavelengths. 
The current linearity solution is most problematic at high fluence levels. When the accumulated signal in a given exposure does not approach saturation, the non-linear response should already be reasonably well-corrected by the current \texttt{NLINFILE}. 
Therefore, while the new linearity solution tends to increase measured source brightness, the effect is lessened at the lower fluence levels that are characteristic of the narrow-band observations taken during WFC3 calibration programs.
\bigbreak
Overall, these data indicate only a $\sim 0.1\%$ increase in the total flux (in electrons) for photometry in a 3-pixel aperture for exposures calibrated with the new linearity solution. 
The RMS variations in the time-dependent IR zeropoints (\cite{Calamida2024}) were $\approx 0.5\%$. 
As such, we deem the image photometry reference file (\texttt{IMPHTTAB}) and filter curves\footnote{Current WFC3/IR filter curves: \texttt{wfc3\_ir\_F????\_008\_syn.fits}} available in CRDS and in use in the automated calibration pipeline are still valid and thus can be used with the new \texttt{NLINFILE} for absolute photometry. 
At this time, new IR zeropoints do not need to be delivered. 

\section*{Conclusion}

The current (2009) linearity reference file provides a quadrant-based average polynomial fit to the IR non-linearity.  Using a new pixel-based non-linearity correction developed by \textcite{shenoy2025}, we test the improvements in recalibrated data for a variety of targets and observing modes, including: internal flat fields, star clusters, imaging, and grism observation of spectrophotometric standard stars used to define the IR zeropoints. 
\textbf{In general, the new pixel-based correction improves linearity at fluence levels above 50,000 $e^- $ (approximately 2/3 of the full well saturation limit), with improvements up to 7\% at the end of the ramp of the \texttt{ima} file for pixels reaching the highest fluence level near 80,000 $e^- $. For 3-pixel aperture photometry in \texttt{flt} images, the typical improvement in photometry is $\sim$0.5\%, where the new solution makes the observed photometry brighter.} 
\bigbreak
Our individual findings for the various test datasets can be summarized as follows:
\begin{enumerate}
    \item Internal Flat Fields: We reprocess internal flat field observations taken with five sample sequence modes which target a range of fluence levels: \texttt{SPARS50}, \texttt{SPARS25}, \texttt{SPARS10}, \texttt{STEP25}, and \texttt{RAPID}. We find the most considerable improvement in \texttt{SPARS25} exposures, where the current non-linearity correction file flags nearly three times as many pixels during the ramp fit (and twice as many pixels for the \texttt{SPARS50} exposure) compared to the new correction. We compute the ratio of the \texttt{ima} instantaneous count rate and \texttt{flt} average count rate as a function of the fluence level ($e^-$) for all internal flat fields in the five sample sequences. We find $\sim0.4$--$5\%$ improvements in the peak-to-peak range over all fluence levels in all sample sequence modes, with the most significant improvement in the \texttt{SPARS25} flats used to compute the coefficients. These results indicate that the new correction decreases the non-linearity across all fluence levels and for a variety of sample sequences.
    
    \item Star Clusters: We compare the count rates in deep exposures of uncrowded regions in two star clusters taken in different sample sequences: NGC 1851 in \texttt{STEP100} and 47 Tucanae in \texttt{SPARS25}. Exposures with many reads (\texttt{NSAMP}=16) were selected to assess changes in the count rate over a range of fluence levels. For NGC 1851, we find that the new correction improves the linearity of a star's peak pixel value by 1--$2\%$ at fluence levels close to saturation. We find a larger improvement of $5\%$ in the linearity of the peak pixel for stars in 47 Tucanae at high fluence levels. Additionally, we find that the \texttt{SPARS25} exposure of 47 Tucanae has a substantially higher fraction of CRHIT flags (DQ=8192) compared to the \texttt{STEP100} exposure of NGC 1851. This suggests that the current 4-sigma rejection threshold used by \texttt{calwf3} to identify cosmic rays during the ramp fitting step may not be appropriate for every sample sequence.
    
    \item Imaging Observations of the G-type Standard Star P330E:  We compare a P330E exposure taken in F110W with both linearity correction applied and find that there is a $\sim3\%$ improvement in the linearity of instantaneous count rates in the last reads before saturation begins with the new pixel-based correction. We analyzed every P330E staring mode exposure in our dataset across three different filters: F110W, F140W, and F160W. We find that in all three filters, the new non-linearity correction file improves the linearity of instantaneous reads at high fluence levels by 2--4\%, with the largest improvement in the F160W exposures.
    
    \item Grism Observations: We compare G102 and G141 grism observations of P330E calibrated with both non-linearity correction files. We find that the scatter in the relative sensitivity of P330E in the visit with the highest exposure time (500 seconds) is reduced in G102 with the new correction. We see the opposite effect in G141; however, this is due to the self-induced persistence in the exposures.

    \item Grism Observation of White Dwarf Standard Star GD71: We plot the (X,Y) position of GD71 observed in grism exposures calibrated with the current and new non-linearity corrections. We compare their normalized sensitivity and find that both corrections return nearly identical normalized sensitivity as a function of position on the detector.
\end{enumerate}

The up-the-ramp fitting step in \texttt{calwf3} uses a 4-$\sigma$ rejection threshold to identify cosmic ray hits for an individual pixel.  This is accomplished by looking for significant deviations from the average in the instantaneous count rates between reads. With the errors in the current linearity file, this results in overly aggressive flagging (as the resulting changes to instantaneous count rates are detected as CRs), where Quadrant 1 (upper left region of the detector) is more susceptible to spurious CR flagging.  While an improvement in photometry is found for all sample sequences, we find that the CR-flagging behavior appears to be related to the sample sequence that was used to acquire the data. For example, \texttt{SPARS} sequences (reads evenly spaced in time) tend to show CR flags populated in early reads of the exposure, whereas \texttt{STEP} sequences (reads evenly spaced early in ramps, then logarithmically spaced later in ramps) tend to have CR flags populated in later reads as the signal approaches full well.  By significantly reducing the number of erroneous CR flags, the total signal used in WFC3/IR ramp fits increases, leading to more precise count rates in the calibrated exposures when using the new correction.
\bigbreak
Finally, we use the new linearity correction to recalibrate all observations of CALSPEC standards used by \textcite{Calamida2024} to quantify any changes to the latest published IR zeropoints. We find that the new solution increases the observed count rate by up to $\sim$0.5\% for F110W observations of G191B2B, for example. However, the weighted average change for all stars in a given filter is only $\sim$0.2\% for bluer filters and $\sim$0.1\% for redder filters and for all narrow-band filters. These changes are smaller than the RMS variations in the published WFC3 zeropoints ($\sim$0.5\%). This suggests that the current image photometry reference file (\texttt{IMPHTTAB}) can still be used together with the new \texttt{NLINFILE} reference file for absolute photometry. 
\bigbreak
The pixel-based \texttt{NLINFILE} \texttt{9au15283i\_lin.fits} was delivered to CRDS and the automated calibration pipeline in October 2025, replacing the 2009 quad-based correction.  This reference file was then superseded by the February 2026 delivery of \texttt{a2412448i\_lin.fits}, incorporating small improvements to the polynomial coefficients discussed in the Addendum below.  Because the latest non-linearity correction affects the observed count rates, we advise users to retrieve updated data products from MAST after all IR data has been reprocessed with \texttt{calwf3} in Spring 2026.

\section*{Recommendations}
CR flags are populated in the DQ array of the \texttt{ima} file and are not propagated into the DQ array of the \texttt{flt} data. These flags can be a useful diagnostic of an error in \texttt{calwf3's} up-the-ramp fit, and we recommend that users compare the \texttt{SCI} and DQ arrays of the \texttt{ima} file to help determine whether the flux in that pixel is reliable in the calibrated \texttt{flt} file. DQ array values $\geq$ 8192, which coincide with the position of sources in the \texttt{SCI}  array or in a particular IR quadrant, can indicate that the ramp fit in \texttt{calwf3} may be inaccurate, affecting the count rates in the \texttt{flt} data products. While the new linearity solution reduces the number of CR flags at high fluence levels, it does not fully resolve CR-flagging issues at the beginning of the ramp, which are more prevalent in \texttt{SPARS} (versus \texttt{STEP}) sample sequences. 
\bigbreak
In general, the new linearity solution significantly reduces non-linearity response of the IR detector. However, the correction still shows 1-2\% deviations from a flat ramp at high fluence levels for some sample sequences (see Figures \ref{fig:intflat_spars50}--\ref{fig:intflat_rapid}). For this reason, we recommend comparing count rates between the  \texttt{ima} and \texttt{flt} data products. For sources with fluence levels exceeding $\sim$2/3 the full well limit ($\sim$50,000 $e^{-}$), the \texttt{flt} photometry may be fainter than in the cumulative read of the \texttt{ima} file (\texttt{[SCI,1]} extension), even with the new linearity correction.
Saturation maps are provided in the \texttt{NODE} extension of the new \texttt{NLINFILE} reference file in units of DN. To convert to electrons, multiply by the mean IR gain of 2.25 $e^{-}/DN$ \parencite{2015gosmeyer}.
\bigbreak
Finally, in some cases, users may wish to mask the last read(s) before sources of interest saturate in long exposures and reprocess the data with \texttt{calwf3} to generate an improved \texttt{flt} file for photometry of bright sources. The process for rejecting reads is described in the following  \href{https://spacetelescope.github.io/hst\_notebooks/notebooks/WFC3/ir\_scattered\_light\_calwf3\_corrections/Correcting\_for\_Scattered\_Light\_in\_IR\_Exposures\_Using\_calwf3\_to\_Mask\_Bad\_Reads.html}{WFC3 Jupyter Notebook}.

\section*{Acknowledgments}
The authors would like to thank Benjamin Kuhn, Joel Green, and Sylvia Baggett for their insightful review of this report. We also like to thank Ralph Bohlin for his expertise and analysis on the IR grism observations of CALSPEC standard stars.

\printbibliography

\section*{Appendix}

\begin{table}[H]
\caption{\textit{Sample sequence, nsamp, exposure time, rootnames, and proposal ID for internal flat field observations used to validate the new linearity correction. All exposures are part of the IR linearity monitoring program taken every year. Sample sequence patterns are listed in the \href{https://hst-docs.stsci.edu/hpiom}{HST Phase II Proposal Instruction Handbook.}}}
\label{tab:intflat}
\begin{tabular}{|c|c|c|>{\raggedright\arraybackslash}p{9cm}|>{\raggedright\arraybackslash}p{2cm}|}
\toprule
Samp\_Seq& n samp & Exptime (s) & Rootname & Proposal ID \\
\midrule
SPARS25 & 16 & 313.122345 & ibmg18gsq, ibmg17s8q, ibmg08vbq, ibmg09h7q & 12352\\
\hline
SPARS25 & 16 & 352.939514 & \shortstack{ibmg01iwq, ibmg16rcq, ibmg15qqq, ibmg14q6q,\\ibmg13gvq, ibmg12z9q, ibmg10uzq, ibmg11vwq,\\ ibmg07jtq,ibmg06edq, ibmg05dfq, ibmg04chq,\\ ibmg03aaq, ibmg02siq, \\ic5n08kkq, ic5n07joq, ic5n09mzq} & \shortstack{12352 \\ [1cm] \\ 13079} \\
\hline
STEP25 & 16 & 274.234863 & ic5n04ahq, ic5n05ysq, ic5n06bmq & 13079 \\
\hline
RAPID & 16 & 43.984360 & ic5n01hwq, ic5n02itq, ic5n03hmq & 13079 \\
\hline
SPARS10 & 16 & 142.945755 & 
\shortstack[t]{ic5n10faq,\\ icfh06ezq, icfh07goq, icfh08hgq, icfh09r3q,\\ icpo04fzq, icpo05ggq, icpo06e7q, icpo07gsq,\\ id0e04drq, id0e05i1q, id0e06k0q, id0e07ebq,\\ idby04adq, idby05tlq, idby06usq, idby07wcq,\\ idou04d2q, idou05n1q, idou06g7q, idou07hyq,\\ idy704deq, idy705g5q, idy706mnq, idy707lrq,\\ ie0j07q6q, ie0j08qgq, ie0j09qoq, ie0j10qyq,\\ iefv07pnq, iefv08rgq, iefv09dyq, iefv10euq,\\ ielx07t2q, ielx08ucq, ielx09hfq, ielx10oyq,\\ iexl07hcq, iexl08myq, iexl09rsq, iexl10zpq,\\ if4y07kuq, if4y08ciq, if4y09l1q, if4y10mcq,\\ ifcn07fsq, ifcn08ggq, ifcn09kiq, ifcn10q8q \\ [0.001cm]} &
\shortstack{13079\\ [0.0012cm] \\ 13563 \\ [0.0012cm] \\ 14009 \\ [0.0012cm] \\ 14375 \\ [0.0012cm] \\ 14538\\ [0.0012cm] \\ 14987 \\ [0.0012cm] \\ 15579\\ [0.0012cm] \\ 15724 \\ [0.0012cm] \\ 16404 \\ [0.0012cm] \\ 16576 \\ [0.0012cm] \\ 17012 \\ [0.0012cm] \\ 17358 \\ [0.0012cm] \\ 17678} \\
\hline
SPARS50 & 11 & 452.936035 & \shortstack[t]{ibvl06snq, ibvl07vtq, ibvl18v7s, ibvl19v4q} & 12696 \\
\bottomrule
\end{tabular}
\end{table}

\begin{longtable}{|c|c|c|c|l|}
\caption{\textit{Proposal ID, filter, sub-array, exposure times, and rootname for P330E observations used to validate the new linearity correction. All exposures were taken in \texttt{RAPID} sample sequence and had n--samps between 5--15 reads. \texttt{ie7h20bvq} is highlighted as it was used to make Figure~\ref{fig:P330E_single}.}} \label{tab:p330etab} \\
\toprule
Proposal ID & Filter & Aperture & Exptime (s) & Rootname \\
\midrule
\endfirsthead
\caption*{\textit{Table~\ref{tab:p330etab} continued}.} \\
\toprule
Proposal ID & Filter & Aperture & Exposure Time (s) & Rootname \\
\midrule
\endhead
\midrule
\multicolumn{5}{r}{Continued on next page} \\
\midrule
\endfoot
\bottomrule
\endlastfoot
11451 & F110W & IRSUB128 & 0.901640 & \shortstack{iaac11gwq, iaac11heq, iaac12tgq,  \\ iaac12tyq, iaac13ieq, iaac13iwq,  \\ iaac14ymq, iaac14z4q}\\
\hline
11451 & F140W & IRSUB128 & 0.901640 & \shortstack{iaac14z6q, iaac14yoq, iaac13iyq,  \\ iaac13igq, iaac12tiq, iaac11hgq,  \\ iaac11gyq, iaac12u0q} \\
\hline
11451 & F160W & IRSUB128 & 0.901640 & \shortstack{iaac12trq, iaac11h7q, iaac12u9q,  \\ iaac13ipq, iaac13j7q, iaac14yxq,  \\ iaac14zfq, iaac11hpq} \\
\hline
11926 & F110W & IRSUB128 & 0.563525 & \shortstack{ibcf60u1q, ibcf60ttq, ibcf61wpq, \\ibcf61wwq} \\
\hline
11926 & F110W & IRSUB128 & 0.788935 & \shortstack{ibcf0ai9q, ibcf0aimq, ibcf0aj8q,  \\ ibcf0ajgq, ibcf0cuzq, ibcf0cvbq} \\
\hline
11926 & F140W & IRSUB128 & 0.788935 & \shortstack{ibcf61wyq, ibcf61wrq, ibcf60u3q,  \\ ibcf60tvq} \\
\hline
11926 & F140W & IRSUB128 & 0.901640 & \shortstack{ibcf0ajiq, ibcf0aibq, ibcf0aioq,  \\ ibcf0aj5q, ibcf0cv1q, ibcf0cvdq} \\
\hline
11926 & F160W & IRSUB128 & 1.352460 & \shortstack{ibcf61wzq, ibcf60u4q, ibcf61wsq,  \\ ibcf60twq} \\
\hline
11926 & F160W & IRSUB128 & 1.577870 & \shortstack{ibcf0ajhq, ibcf0cv3q, ibcf0aj3q,  \\ ibcf0aiqq, ibcf0aidq, ibcf0cvfq} \\
\hline
11926 & F160W & IRSUB64 & 0.243096 & \shortstack{ibcf23rtq, ibcf24oxq, ibcf23s9q,  \\ ibcf22grq, ibcf22g9q, ibcf21ynq,  \\ ibcf21xgq, ibcf24pdq} \\
\hline
12334 & F110W & IRSUB128 & 0.450820 & \shortstack{ibnx16h9q, ibnx16h3q, ibnx16gyq,  \\ ibnx04dcq, ibnx04d6q, ibnx04d1q} \\
\hline
12334 & F140W & IRSUB128 & 0.676230 & \shortstack{ibnx16hbq, ibnx16h5q, ibnx04d8q,  \\ ibnx04deq, ibnx04d3q, ibnx16h0q} \\
\hline
12334 & F160W & IRSUB128 & 1.239755 & \shortstack{ibnx16hcq, ibnx16h6q, ibnx16h1q,  \\ ibnx04dfq, ibnx04d9q, ibnx04d4q} \\
\hline
12699 & F110W & IRSUB128 & 0.450820 & ibwi02miq, ibwi02mnq, ibwi02msq \\
\hline
12699 & F140W & IRSUB128 & 0.676230 & ibwi02mkq, ibwi02mpq, ibwi02muq \\
\hline
12699 & F160W & IRSUB128 & 1.239755 & ibwi02mlq, ibwi02mvq, ibwi02mqq \\
\hline
13089 & F140W & IRSUB128 & 0.676230 & \shortstack{ic5s14jrq, ic5s14jjq, ic5s10hjq,  \\ ic5s10hbq} \\
\hline
13089 & F160W & IRSUB128 & 1.014345 & \shortstack{ic5s10hcq, ic5s14jkq, ic5s10hkq,  \\ ic5s14jsq} \\
\hline
13573 & F160W & IRSUB128-FIX & 1.577870 & icfs12ajq, icfs12akq \\
\hline
13573 & F160W & IRSUB256-FIX & 3.889411 & icfs12alq, icfs12amq \\
\hline
13575 & F110W & IRSUB128 & 0.450820 & \shortstack{ich319ngq, ich319naq, ich319n4q,  \\ ich319nmq} \\
\hline
13575 & F140W & IRSUB128 & 0.563525 & \shortstack{ich319npq, ich319njq, ich319n7q,  \\ ich319ndq} \\
\hline
13575 & F160W & IRSUB128 & 1.014345 & \shortstack{ich319n8q, ich319neq, ich319nkq,  \\ ich319nqq} \\
\hline
14021 & F110W & IRSUB64 & 0.243096 & icrw02eoq, icrw02enq \\
\hline
14021 & F140W & IRSUB64 & 0.425418 & icrw02elq, icrw02emq \\
\hline
14021 & F160W & IRSUB128 & 0.901640 & icrw02evq, icrw02ewq \\
\hline
14384 & F110W & IRSUB64 & 0.243096 & id2402a5q, id2402a4q \\
\hline
14384 & F140W & IRSUB128 & 0.450820 & id2402a7q, id2402a6q \\
\hline
14384 & F160W & IRSUB128 & 0.901640 & id2402afq, id2402aeq \\
\hline
14883 & F110W & IRSUB64 & 0.243096 & iddn09jwq, iddn09jxq \\
\hline
14883 & F140W & IRSUB128 & 0.450820 & iddn09jyq, iddn09jzq \\
\hline
14883 & F160W & IRSUB128 & 0.901640 & iddn09k6q, iddn09k7q \\
\hline
14992 & F110W & IRSUB64 & 0.243096 & idor09n3q, idor09n2q \\
\hline
14992 & F140W & IRSUB128 & 0.450820 & idor09n5q, idor09n4q \\
\hline
14992 & F160W & IRSUB128 & 0.901640 & idor09ncq, idor09ndq \\
\hline
15582 & F110W & IRSUB64 & 0.243096 & idw109oqq, idw109opq \\
\hline
15582 & F140W & IRSUB128 & 0.450820 & idw109orq, idw109osq \\
\hline
15582 & F160W & IRSUB128 & 0.901640 & idw109ozq, idw109p0q \\
\hline
16030 & F110W & IRSUB64 & 0.425418 & \textbf{ie7h20bvq}, ie7h20bwq \\
\hline
16030 & F140W & IRSUB128 & 0.788935 & ie7h20bxq, ie7h20byq \\
\hline
16030 & F160W & IRSUB128 & 1.127050 & ie7h20c6q, ie7h20c5q \\
\hline
16415 & F110W & IRSUB64 & 0.425418 & iege22rlq, iege22rmq \\
\hline
16415 & F140W & IRSUB128 & 0.788935 & iege22roq, iege22rnq \\
\hline
16415 & F160W & IRSUB128 & 1.127050 & iege22rzq, iege22s0q \\
\hline
16579 & F110W & IRSUB64 & 0.425418 & iemg20b7q, iemg20b6q \\
\hline
16579 & F140W & IRSUB128 & 0.788935 & iemg20b8q, iemg20b9q \\
\hline
16579 & F160W & IRSUB128 & 1.127050 & iemg20bhq, iemg20bgq \\
\hline
17015 & F110W & IRSUB64 & 0.425418 & iev520xpq, iev520xoq \\
\hline
17015 & F140W & IRSUB128 & 0.788935 & iev520xqq, iev520xrq \\
\hline
17015 & F160W & IRSUB128 & 1.127050 & iev520xyq, iev520xzq \\
\hline
\end{longtable}

\begin{table}[H]
\centering
\caption{\textit{Grism, date (year), rootname, proposal IDs, exposure times, and sample sequence of grism observations of P330E used to validate the new linearity correction.}}
\label{tab:p330egrism}
\begin{tabular}{|c|c|c|c|c|c|}
\toprule
Grism & Date (Year) & Rootname & Proposal ID & Exptime (s) & Samp\_Seq \\
\midrule
G102 & 2010.51 & ibcf0cvnq & 11926 & 43.98 & RAPID \\
G102 & 2010.71 & ibcf61xnq & 11926 & 13.83 & STEP25 \\
G102 & 2012.75 & ibwib6m8q & 12699 & 499.23 & STEP50 \\
G102 & 2012.75 & ibwib6m9q & 12699 & 499.23 & STEP50 \\
G102 & 2012.75 & ibwib6mbq & 12699 & 499.23 & STEP50 \\
G102 & 2012.75 & ibwib6mdq & 12699 & 499.23 & STEP50 \\
G102 & 2012.75 & ibwib6mfq & 12699 & 499.23 & STEP50 \\
G102 & 2014.08 & ich319ogq & 13575 & 72.94 & SPARS10 \\
G102 & 2025.21 & ifdg04a2q & 17681 & 32.25 & RAPID \\
G102 & 2025.21 & ifdg04a4q & 17688 & 32.25 & RAPID \\
G102 & 2025.21 & ifdg04a7q & 17688 & 32.25 & RAPID \\
G102 & 2025.23 & ifi520f6q & 17688 & 32.25 & RAPID \\
G102 & 2025.25 & ifdg05awq & 17688 & 32.25 & RAPID \\
\hline
\hline
G141 & 2010.51 & ibcf0cvoq & 11926 & 32.25 & RAPID \\
G141 & 2010.71 & ibcf61xoq & 11926 & 13.83 & STEP25 \\
G141 & 2012.61 & ibtwb4bcq & 12336 & 23.45 & RAPID \\
G141 & 2012.61 & ibtwb4bdq & 12336 & 102.94 & SPARS10 \\
G141 & 2012.61 & ibtwb4bfq & 12336 & 23.45 & RAPID \\
G141 & 2012.61 & ibtwb4bgq & 12336 & 102.94 & SPARS10 \\
G141 & 2012.61 & ibtwb4biq & 12336 & 23.45 & RAPID \\
G141 & 2012.61 & ibtwb4bjq & 12336 & 102.94 & SPARS10 \\
G141 & 2012.61 & ibtwb4blq & 12336 & 23.45 & RAPID \\
G141 & 2012.61 & ibtwb4bmq & 12336 & 102.94 & SPARS10 \\
G141 & 2012.61 & ibtwb4boq & 12336 & 23.45 & RAPID \\
G141 & 2012.61 & ibtwb4bpq & 12336 & 102.94 & SPARS10 \\
G141 & 2014.08 & ich319oeq & 13575 & 43.98 & RAPID \\
G141 & 2025.21 & ifdg04a9q & 17688 & 23.45 & RAPID \\
G141 & 2025.21 & ifdg04acq & 17688 & 23.45 & RAPID \\
G141 & 2025.21 & ifdg04aeq & 17688 & 23.45 & RAPID \\
G141 & 2025.23 & ifi520f4q & 17681 & 23.45 & RAPID \\
\bottomrule
\end{tabular}
\end{table}

\begin{table}[H]
\centering
\caption{\textit{Grism filter and rootname for every GD71 exposure used to validate the new linearity correction. All exposures are 103 seconds long, taken in \texttt{SPARS25}, and are a part of WFC3 program 14024. Bolded exposures are the exposure that the data is normalized to in Figure \ref{fig:gd71pos}.}}
\label{tab:gd71pos}
\begin{tabular}{|c|c||c|c|}
\toprule
Grism & Rootname & Grism & Rootname\\ 
\midrule
G102 & \textbf{icqw01z1q} & G141 & \textbf{icqw02h3q}\\
G102 & icqw01zjq & G141 & icqw02haq\\
G102 & icqw01zrq & G141 & icqw02haq\\
G102 & icqw01zvq & G141 & icqw02i2q\\
G102 & icqw01zyq & G141 & icqw02i5q\\
G102 & icqw01auq & G141 & icqw02i9q\\
G102 & icqw01ayq & G141 & icqw02icq\\
G102 & icqw01b1q & G141 & icqw02igq\\
G102 & icqw01b5q & G141 & icqw02ijq\\
G102 & icqw01b9q & G141 & icqw02inq\\
G102 & icqw01bcq & G141 & icqw02iqq\\
\bottomrule
\end{tabular}
\end{table}

\vspace{22pt}
\section*{Addendum: Updated IR Non-Linearity Coefficients}
\setcounter{figure}{0} 
\renewcommand{\thefigure}{A\arabic{figure}} 

\begin{figure}[H]
\begin{center}
\includegraphics[width=\linewidth]{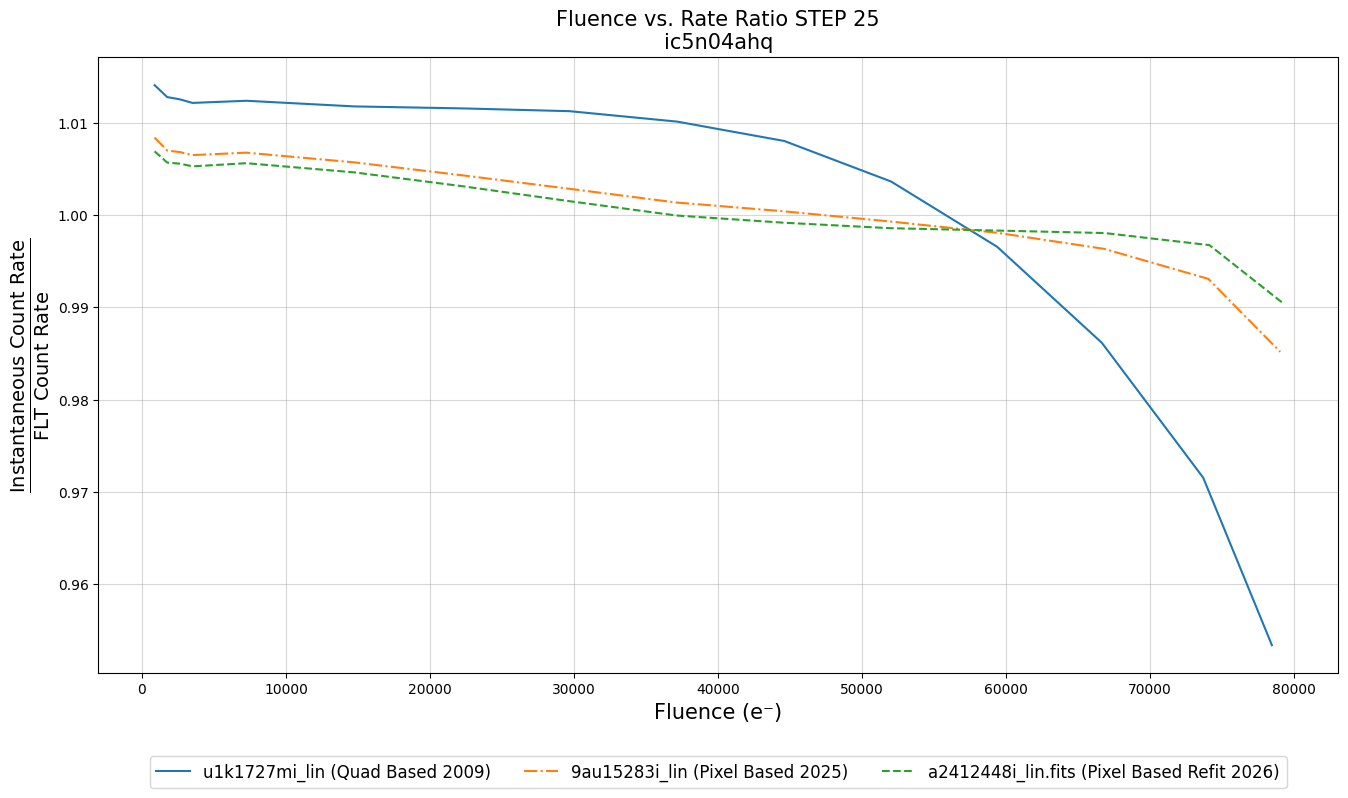}
\end{center}
\caption{\textit{Similar to Figure \ref{fig:intflat_step25}, but now showing the improvement in the ramp for a representative \texttt{STEP25} INTFLAT exposure calibrated with the 2026 pixel-based non-linearity reference file (green), the 2025 pixel-based solution (orange), and the 2009 quad-based solution (blue).}
}
\label{fig:intflat_step25_prepost}
\end{figure}

The \texttt{calwf3} implementation of the IR non-linearity correction is a multiplicative increase of signal that takes the form $(1+\delta)$, where $\delta$ is a third-order polynomial of the measured fluence (see \S3.3.6 of the \textit{WFC3 Data Handbook} \parencite{DHB}).  This convention matches that adopted for the \textit{JWST} NIRCam detector (Eq.\ 5 of \textcite{2017Canipe}) and its corresponding reduction pipeline.  
\bigbreak
Following historical convention (\textit{eg}., Eq.\ 4 of \textcite{2014Hilbert}), the 2025 pixel-based IR non-linearity fitting determined the third-order polynomial approximation of the fluence \textit{deficit} (Eq.\ 3 of \textcite{shenoy2025}).  Rather than modifying \texttt{calwf3} to divide the measured signal by this deficit, the WFC3 Team approximated the multiplicative correction by negating all the polynomial coefficients -- essentially making the substitution $(1+\delta) \approx (1-\delta)^{-1}$ (see Eqs.\ 2 and 3 of \textcite{shenoy2025}).  This substitution is acceptable when $\delta \ll 1$. However the WFC3 Team subsequently concluded that IR non-linearity becomes large enough at the highest (non-saturated) fluence levels to significantly ($>\!0.3$\%) compromise this approximation.  
\bigbreak
As of February 2026, the WFC3 Team has swapped in a revised set of coefficients, now based on a third-order polynomial fit to the reciprocal of the $(1-\delta)$ polynomial expansion (rather than just negating the polynomial coefficients).  \textbf{The updated NLINFILE \texttt{a2412448i\_lin.fits} will be used by the \texttt{calwf3} pipeline for reprocessing all IR observations in  Spring 2026.}  The WFC3 Team has extensively tested the impact of the latest revised IR non-linearity coefficients, as summarized in the plots and discussion to follow.  
\bigbreak
Figure \ref{fig:intflat_step25_prepost} shows the effect of the revised non-linearity coefficients for the case of a single \texttt{STEP25} INTFLAT selected from Figure \ref{fig:intflat_step25}.  The updated coefficients are seen to improve linearity of the flat field ramp at $\approx$ all fluence levels (\textit{green curve}) compared to the ramp based on the 2025 coefficients (\textit{orange curve}), and most noticeably for the highest fluence levels.

\begin{figure}[H]
\begin{center}
\includegraphics[width=\linewidth]{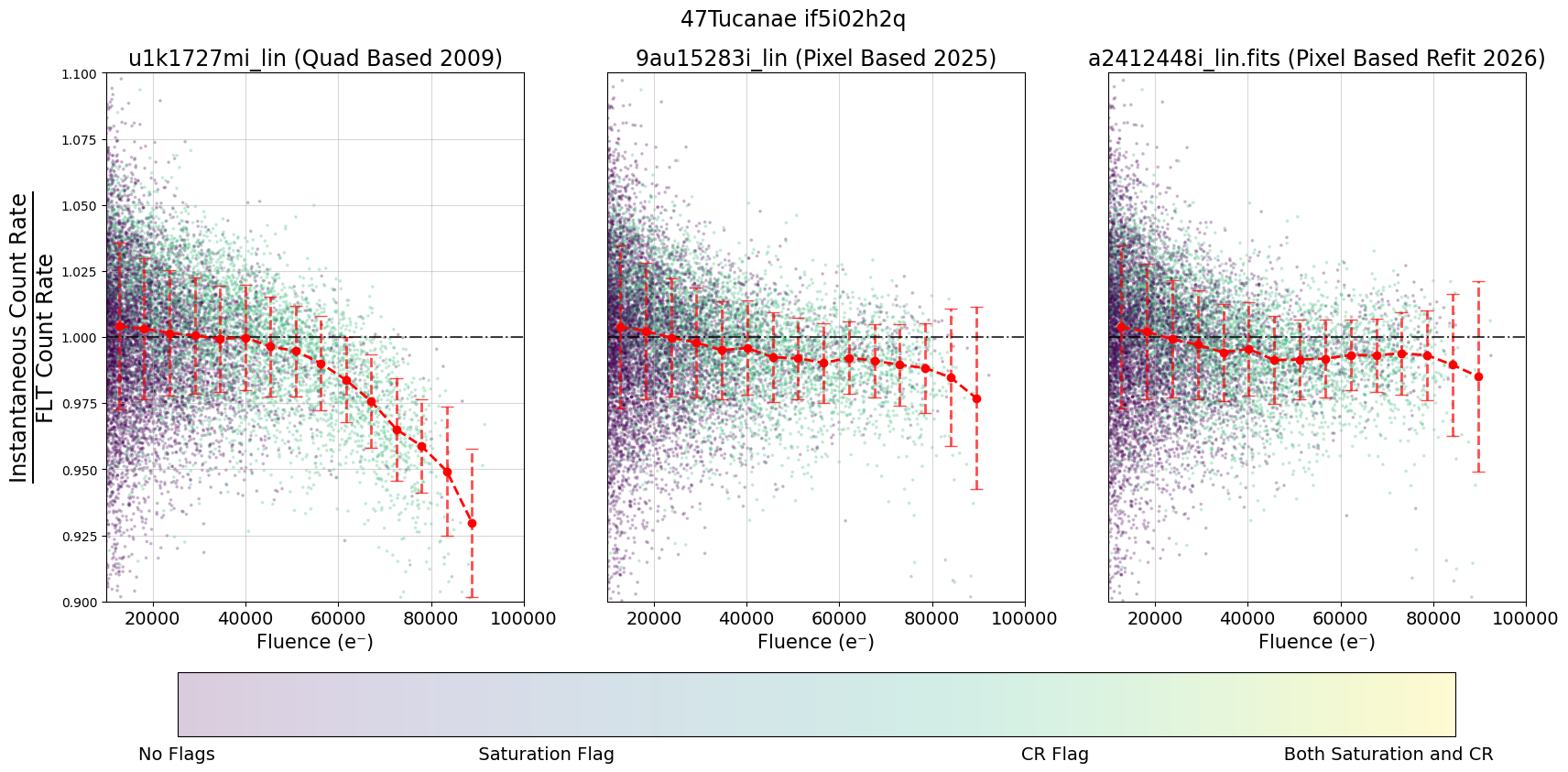}
\end{center}
\caption{\textit{Similar to Figure \ref{fig:47tuc_instcr}, but now additionally showing the improvement in the ramp for an image of 47 Tucanae calibrated with the 2026 pixel-based non-linearity reference file (right), compared to the results using the 2025 pixel-based reference file (middle) and the 2009 quad-based solution (left).}
}
\label{fig:47tuc_prepost_3panel}
\end{figure}

Additional test results using the 2026 revised coefficients are shown in the right panel of Figure \ref{fig:47tuc_prepost_3panel} for the 47 Tucanae data presented in Figure \ref{fig:47tuc_instcr}. Small improvements can be seen in the high fluence regime, similar to the results found for \texttt{INTFLAT} data. 

Lastly, the WFC3 Team has tested the impact of the 2026 non-linearity solution on the IR filter zeropoints.  In all cases, changes in the mean flux ratio for the five CALSPEC photometric standards are found to be negligible ($<<\!0.1$\%) compared to the 2025 solution (M. Marinelli 2025, priv.\ comm.).  This is unsurprising, given that the peak fluence levels of the photometric standard stars fall in a region of Figure \ref{fig:47tuc_prepost_3panel} ($\sim\!60000$ e${}^-$) showing minimal change in the observed count rates due to the updated coefficients.
\end{document}

%% file: obsobs_p330e.tex
\begin{figure}[H]
    \centering
    \includegraphics[width=\linewidth]{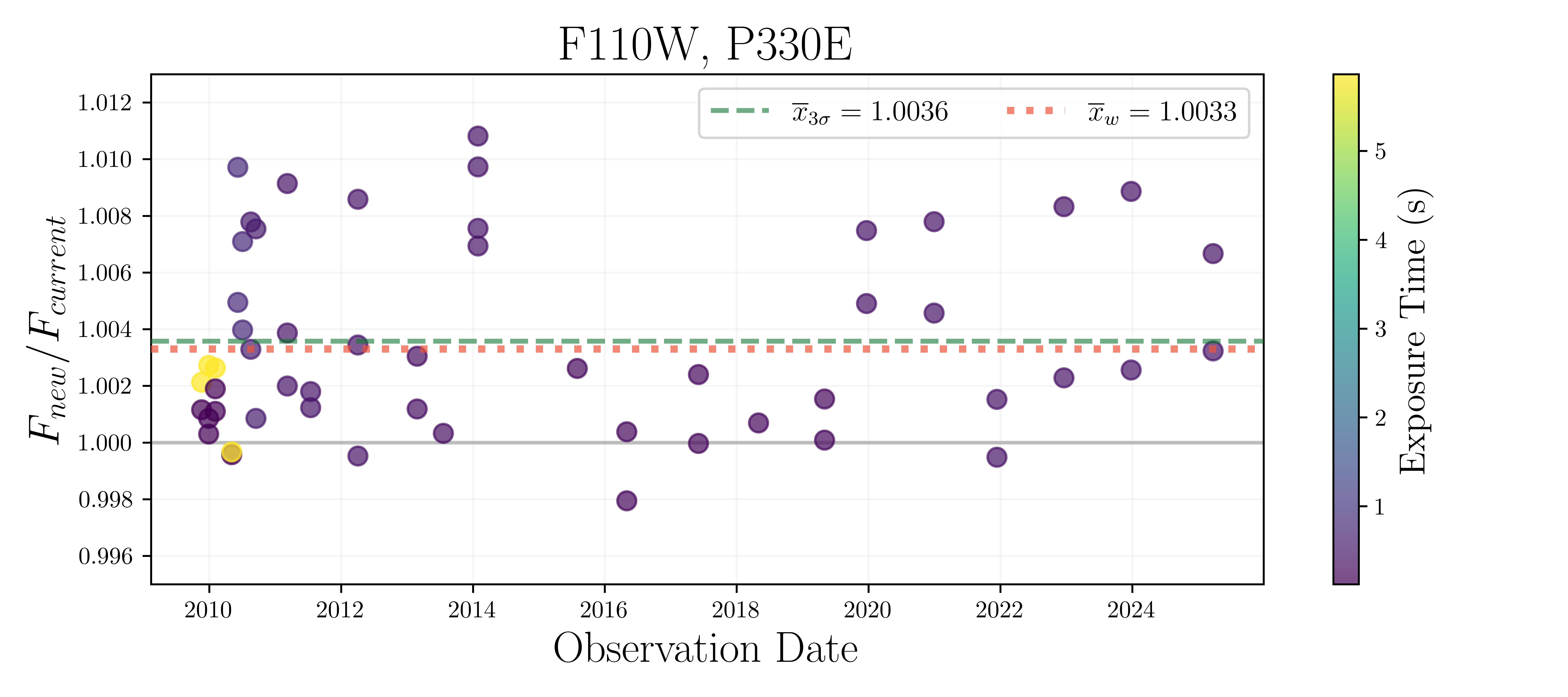}
    \caption{\textit{Flux ratios as a function of time for \texttt{flt} observations of P330E in the F110W filter, comparing the measured flux ($\,e^{-}/s\,$) after implementing the new linearity solution ($\,F_{new}\,$) to the flux using the current (2009) calibration ($\,F_{current}\,$). Points are colored by exposure time ($s$), the $3\sigma$-clipped mean is a green dashed line, and the mean flux ratio (weighted by photometric errors) is a pink dotted line.}}
    \label{fig:obsobs_p330e}
\end{figure}

%% file: obsobs_gd153.tex
\begin{figure}[H]
    \centering
    \includegraphics[width=\linewidth]{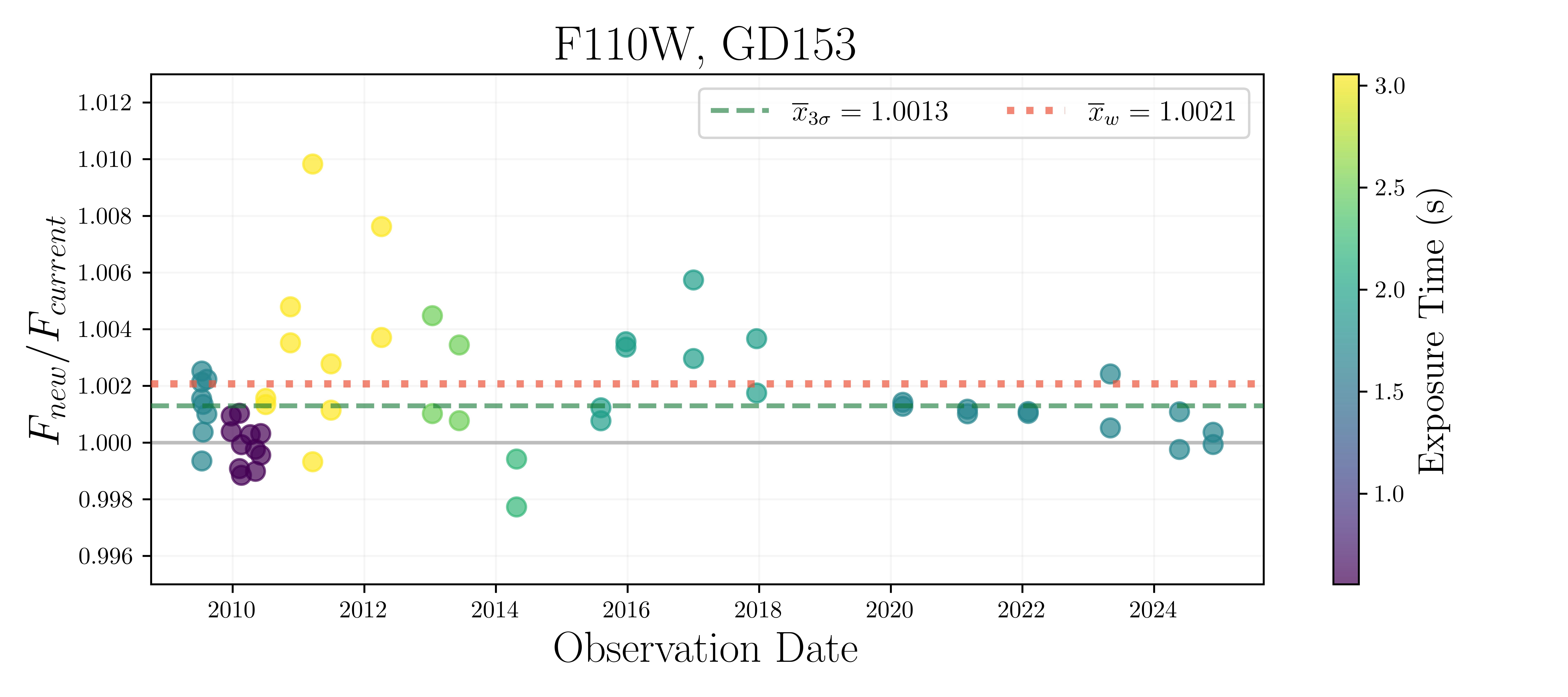}
    \caption{\textit{Same as Figure \ref{fig:obsobs_p330e}, but for GD-153.}}
    \label{fig:obsobs_gd153}
\end{figure}

%% file: obsobs.tex
\begin{figure}[H]
    \centering
    \includegraphics[width=\linewidth]{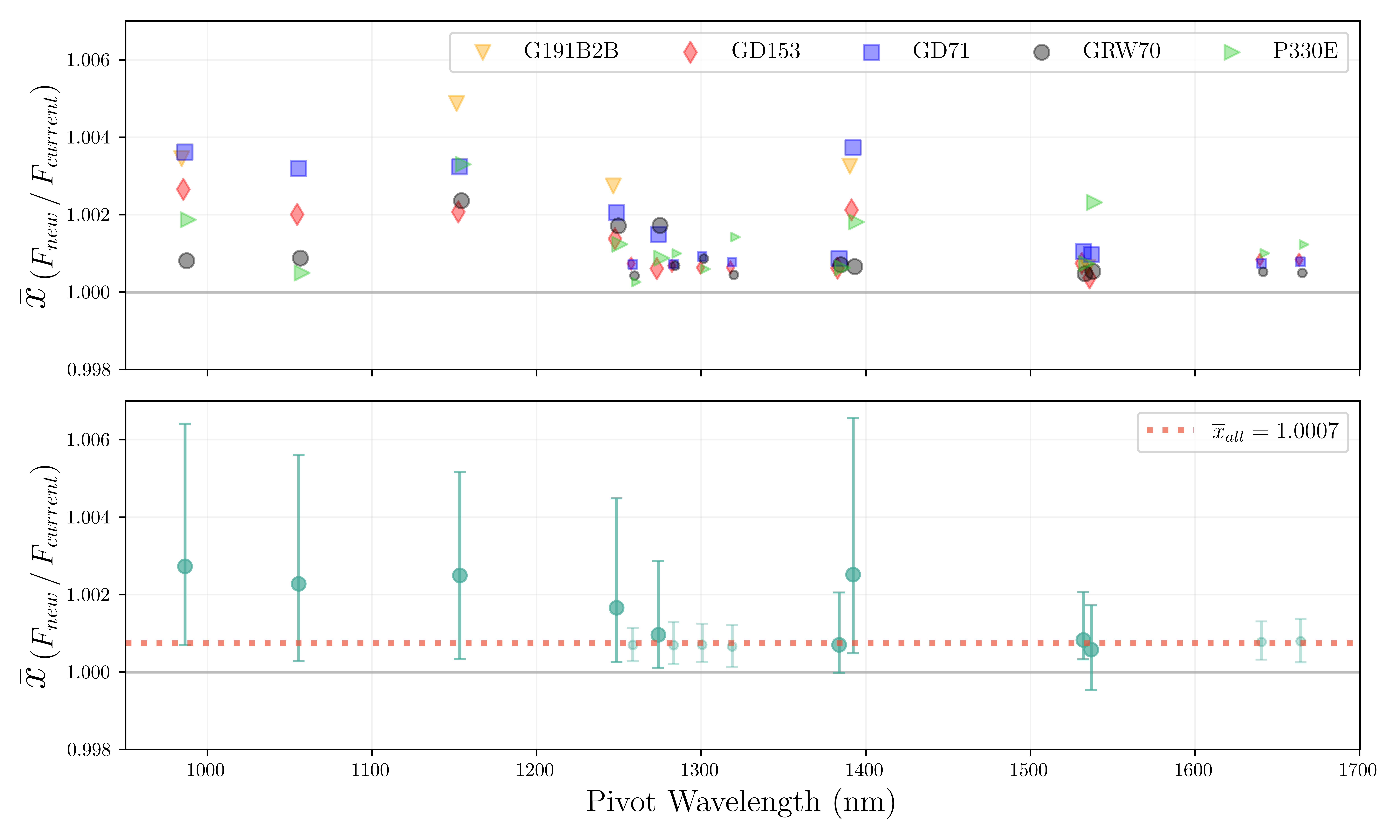}
    \caption{\textit{Mean flux ratios (observed count rates for \texttt{flt} data calibrated with the new linearity solution, divided by the observed count rates for the current (2009) solution), weighed by photometric errors, as a function of filter pivot wavelength. Narrow filters are indicated with a smaller marker size. Top: Mean flux ratio per target/filter combination. Bottom: Mean flux ratio for all stars within each filter. Error bars show the median absolute deviation above and below the mean.  
  }}
    \label{fig:obs_percentdiff}
\end{figure}
